\documentclass{aa}

\usepackage[dvipsnames]{xcolor} 
\usepackage{graphicx, subfigure, amsmath, pifont}
\usepackage{amssymb}
\usepackage{multicol}
\usepackage[breaklinks=true]{hyperref} 
\hypersetup{colorlinks=true,citecolor=Blue}
\usepackage{natbib}
\bibpunct{(}{)}{;}{a}{}{,} % to follow the A&A style
\usepackage[english]{babel}
\usepackage{blindtext}
\usepackage[normalem]{ulem}
\usepackage{comment}
\usepackage{lmodern}
\usepackage{ulem} 
\usepackage{soul}
\usepackage{lscape}
\usepackage{longtable}
\usepackage{soul}
\usepackage{lipsum}
\usepackage{hyperref}
\usepackage{chngcntr}
\usepackage{txfonts}

\let\oldpageref\pageref
\renewcommand{\pageref}{\oldpageref*}

\let\linenumbers\relax

\graphicspath{{./}{figures/}}

\begin{document}

\title{MINDS. Young binary systems with JWST/MIRI: \\variable water-rich primaries and extended emission}
\titlerunning{MINDS. Binaries with JWST-MIRI}

\author{
Nicol\'as T. Kurtovic\orcid{0000-0002-2358-4796}\inst{1} 
\and Sierra L. Grant\orcid{0000-0002-4022-4899}\inst{2}
\and Milou Temmink\orcid{0000-0002-7935-7445}\inst{3}
\and Andrew D. Sellek \orcid{0000-0003-0330-1506}\inst{3}
\and Ewine F. van Dishoeck\orcid{0000-0001-7591-1907}\inst{3,1}
\and Thomas Henning\orcid{0000-0002-1493-300X}\inst{4} 
\and Inga Kamp\orcid{0000-0001-7455-5349}\inst{5} 
\and Valentin Christiaens\orcid{0000-0002-0101-8814}\inst{6, 7}
\and Andrea Banzatti\orcid{0000-0003-4335-0900}\inst{8}
\and Danny Gasman\orcid{0000-0002-1257-7742}\inst{6}
\and Till Kaeufer\orcid{0000-0001-8240-978X}\inst{9}
\and Lucas M. Stapper\orcid{0000-0001-9524-3408}\inst{4} 
\and Riccardo Franceschi\orcid{0000-0002-8889-2992}\inst{4}
\and Manuel G\"udel\orcid{0000-0001-9818-0588}\inst{10,11}
\and Pierre-Olivier Lagage\inst{12}
\and Marissa Vlasblom\orcid{0000-0002-3135-2477}\inst{3}
\and Giulia Perotti\orcid{0000-0002-8545-6175}\inst{13, 4}
\and Kamber Schwarz\orcid{0000-0002-6429-9457}\inst{4}
\and Alice Somigliana\orcid{0000-0003-2090-2928}\inst{4}
}

\institute{
Max-Planck-Institut f\"ur Extraterrestrische Physik, Giessenbachstrasse 1, D-85748 Garching, Germany\label{mpe}, \\\email{kurtovic@mpe.mpg.de} 
\and Earth and Planets Laboratory, Carnegie Institution for Science, 5241 Broad Branch Road, NW, Washington, DC 20015, USA
\and Leiden Observatory, Leiden University, P.O. Box 9513, 2300 RA Leiden, the Netherlands 
\and Max-Planck-Institut f\"{u}r Astronomie (MPIA), K\"{o}nigstuhl 17, 69117 Heidelberg, Germany
\and Kapteyn Astronomical Institute, Rijksuniversiteit Groningen, Postbus 800, 9700AV Groningen, The Netherlands
\and Institute of Astronomy, KU Leuven, Celestijnenlaan 200D, 3001 Leuven, Belgium
\and STAR Institute, Universit\'e de Li\`ege, All\'ee du Six Ao\^ut 19c, 4000 Li\`ege, Belgium
\and Department of Physics, Texas State University, 749 North Comanche Street, San Marcos, TX 78666, USA 
\and School of Physics and Astronomy, University of Exeter, Stocker Road, Exeter EX4 4QL, UK
\and Dept. of Astrophysics, University of Vienna, T\"urkenschanzstr. 17, A-1180 Vienna, Austria
\and ETH Z\"urich, Institute for Particle Physics and Astrophysics, Wolfgang-Pauli-Str. 27, 8093 Z\"urich, Switzerland
\and Université Paris-Saclay, Université Paris cité, CEA, CNRS, AIM, F-91191 Gif-sur-Yvette, France.
\and Niels Bohr Institute, University of Copenhagen, NBB BA2, Jagtvej 155A, 2200 Copenhagen, Denmark
}

\date{Accepted 28 July 2025}

% F.M. 0000-0002-1637-7393

  \authorrunning{N.T.~Kurtovic}
  \titlerunning{Binaries with JWST/MIRI-MRS}
% \abstract{}{}{}{}{} 
% 5 {} token are mandatory

  \abstract
  % context heading (optional)
    {Dynamical disk-companion interactions can have a large impact on the evolution of circumstellar disks, as these can produce perturbations to the material distribution, density, and temperature, affecting their potential for planet formation. } 
  % aims heading (mandatory)
    {As part of the JWST GTO program MINDS, we analyze the mid-infrared emission of three Class II binary systems: VW\,Cha, WX\,Cha, and RW\,Aur, to investigate the impact of stellar multiplicity on the chemistry and physics of their inner disk. }
  %methods heading (mandatory)
    {We analyze the 1D spectrum from JWST/MIRI-MRS for primary and secondary disks separately, extracted by combining forward modeling with a theoretical PSF and aperture photometry. After continuum subtraction, we modeled the molecular lines with 0D slab models. We interpret the results by comparing our JWST spectra to VLT/CRIRES+, Spitzer/IRS. The extended mid-infrared emission is compared to ALMA, for which we also include the binary DF\,Tau in our sample. } 
  % results heading (mandatory)
    {Primary and secondary disks are dramatically different in their mid-infrared emission, with primary disks showing H$_2$O-rich spectra, and secondary disks being mostly line poor to the sensitivity of our spectra. When comparing MIRI-MRS to Spitzer/IRS, we observe large variability in the line emission of VW\,Cha\,A, as well as in the continuum of RW\,Aur\,A. The disks around VW\,Cha\,BC and RW\,Aur\,B show evidence of ionizing radiation, and a further comparison with ALMA at high angular resolution dust continuum suggest that the spectrum of RW\,Aur\,B is well explained by its $\sim 4$\,au cavity. 
    All the systems show [Ne II] jet emission, and three of them also show spatially resolved emission structures in H$_2$, likely originated by outflows and dynamical interactions. } 
  % conclusions heading (optional), leave it empty if necessary  {}
    %{}
    {Many of the observed features in the primary disks, such as enhanced water emission, could be linked to the increased accretion and radial drift produced by dynamical disk truncation. However, additional mechanisms are needed to explain the large differences between primary and secondary disks, potentially inner disk substructures. This work is an example of the need for combining multiple facilities to fully understand the observations from JWST. }


   \maketitle

\section{Introduction}

Most of the stars are formed in a binary or higher multiplicity stellar systems \citep[see][]{Duchene2013, offner2023}. These external companions can have a significant impact on the planet formation potential of circumstellar disks, as dynamical disk-companion interactions can truncate the outer disk radii, warp inner disk regions, and launch material into eccentric or unbound orbits \citep[e.g., ][]{papaloizou1977, artymowicz1994, Dai2015, manara2019, cuello2020, rota2022, Rowther2022, zagaria2023}. Over the last decade, high-angular resolution facilities such as the Atacama Large (sub-)Millimeter Array (ALMA) and the Very Large Telescope (VLT/SPHERE) have found direct observational evidence of such dynamical interactions between young stellar objects (YSOs), and their impact on the material distribution of the outer disk \citep{Cabrit2006, mayama2010, fernandez2017, kurtovic2018, kepler2020, menard2020, Zapata2020, Dong2022, Weber2023}. With the growing number of systems observed by the Mid-InfraRed Instrument \citep[MIRI][]{rieke2015, wright2015, wright2023} of the James Webb Space Telescope \citep[JWST,][]{rigby2023}  , we can now also explore the impact of stellar multiplicity in the inner disk chemistry of each star.

Among the numerous findings with the Medium Resolution Spectrometer mode of JWST/MIRI (MRS, 4.9-27\,$\mu$m, \citet{argyriou2023}), studies have revealed a large diversity of molecular emission lines, and particularly on the C/O ratio, ranging from spectra dominated by complex carbon-species \citep[e.g.,][]{tabone2023, kanwar2024, arabhavi2024, colmenares2024, long2025}, to disks dominated by H$_2$O emission lines \citep[e.g.,][]{gasman2023, perotti2023, xie2023, temmink2024, grant2024, romeromirza2024, pontoppidan2024, banzatti2025, gasman2025}, including disks where H$_2$O and other species such as HCN, C$_2$H$_2$, and CO$_2$ show a more comparable peak brightness \citep[][]{grant2023, vlasblom2025}. 
This diversity has been connected to the outer disk evolution and the drift of icy pebbles from the outer disk \citep[e.g.,][]{mah2023, mah2024}, which feeds the inner disk with volatiles through ice sublimation. Thus, for disks whose evolution is dominated by the radial drift of pebbles \citep[see][]{trapman2019}, the inner disk composition should show different signatures when compared to another disk where radial drift has been halted by strong dust traps \citep[see ][]{banzatti2020, kalyaan2021, banzatti2023b, kalyaan2023, gasman2025, sellek2025}. 
In this scenario, the circumstellar disks in binary systems could provide a test-case to study the influence of drift on the inner disk mid-infrared emission, as dynamical disk truncation from the companions will perturb the material into the inner regions \citep{zagaria2023}, thus replenishing the inner disk with material from the outer disk. 

The dynamical influence of an external companion will also have a time dependency, as it is more pronounced when the stars are at periastron, and weaker for larger distances. Thus, stellar systems where the binary orbit is known provide an ideal test for time-variable signatures in the inner disk chemistry, and set constraints over heating and cooling processes. 

The first analysis of a multiple stellar system with JWST/MIRI-MRS was done for the DF\,Tau system \citep{grant2024}, revealing a line forest associated to water emission at different temperatures, making it one of the richest T-Tauri disk spectrum observed to date. With a separation of about 10\,au in the sky plane ($\approx70$\,mas), JWST/MIRI-MRS could only recover the combined spectra of these binaries, relying on the combination of several additional instruments to interpret the mid-infrared signatures. \citet{grant2024} showed the importance of having complementary data to interpret JWST observations, but the small separation of the binaries was a limiting factor to quantify the contribution of each disk to the MIRI-MRS spectra.

As part of the MIRI mid-INfrared Disk Survey (MINDS) JWST guaranteed time observation program \citep[PID: 1282, PI: T. Henning,][]{kamp2023, henning2024}, we present new JWST/MIRI-MRS observations of three known Class II wide binary systems: VW\,Cha, WX\,Cha, and RW\,Aur, which we describe in detail in Sect.~\ref{sec:intro:systems}. In contrast to DF\,Tau, these three systems have binary separations larger than the angular resolution of JWST/MIRI-MRS at its shorter wavelengths (separations in the range 0.7''-1.5'', or 100-240\,au at the distance of these sources). The main focus of our methodology is to disentangle the emission from each binary component and obtain the spectrum of each disk, as described in Sect.~\ref{sec:obs:data}. To aid in the interpretation of the spectra, we also searched for extended emission structures which could be compared to other observational facilities. An analysis of the 1D spectra, including the properties of the atomic and molecular emission lines, is presented in Sect.~\ref{sec:res:slabs}, while the analysis of the extended emission is presented in Sect.~\ref{sec:ext:extended}. We discuss the interpretation of our findings in Sect.~\ref{sec:disc}, and summarize our conclusions in Sect.~\ref{sec:conclusions}.

\section{Multiple Stellar Systems} \label{sec:intro:systems}

This work includes new observations of VW\,Cha, WX\,Cha, and RW\,Aur with JWST/MIRI-MRS. For comparison, we also include the observation of the binary DF\,Tau, which was previously analyzed in \citet{grant2024}. This shorter period binary provides a point of comparison for our moderate-separation systems. In addition, we also show for the first time the extended gas emission of DF\,Tau as detected with ALMA and JWST/MIRI-MRS. These binaries are close to being equal mass companions (mass ratio $q\approx1$), with the exception of WX\,Cha, where $q>2$.

\subsection{VW Cha}

VW\,Cha is a quadruple stellar system located at 185\,pc \citep{gaia2021edr3}, with VW\,Cha\,A being a spectroscopic binary \citep{melo2003, nguyen2012} of a tentative 10\,days period \citep{zsidi2022}. This spectroscopic binary is consistent with stellar templates between K5 to K7 \citep{zsidi2022}, and at least one of these stars is consistent with a mass of M$_\star=0.7_{-0.35}^{+0.50}$\,M$_\odot$ \citep{daemgen2013}. Located at about 0.7'' (130\,au in the sky-plane) from VW\,Cha\,A is the pair VW\,Cha\,B and C, which are separated by 0.1'' from each other (19\,au in the sky-plane) \citep{brandeker2001, Correia2006, vogt2012, daemgen2013}. The combined optical spectrum of VW\,Cha\,BC is consistent with spectral types between M0 to M2.5, and at least one of them with mass M$_\star=0.57_{-0.19}^{+0.28}$\,M$_\odot$ \citep{daemgen2013}. In a multi-filter photometric campaign by \citet{zsidi2022}, VW\,Cha\,A was demonstrated to have an optical brightness variability of up to 0.8\,mag over the span of one day, with evidence of accretion luminosity changing from 0.8\,L$_\odot$ to 2.3\,L$_\odot$ in three days, as well as a low velocity outflow detected in the [O I] optical forbidden line, indicative of a wide angle wind \citep{zsidi2022}. Due to the lack of studies at high angular resolution at millimeter wavelengths, the gas and dust distribution of the outer disks remains unconstrained.

\subsection{WX Cha}

WX\,Cha is a binary system located at 191\,pc \citep{gaia2021edr3}. The primary star of the system, WX\,Cha\,A, is an M0 star \citep{fiorellino2022} of mass $0.49\pm0.12$\,M$_\odot$ \citep{daemgen2013}. Located at 0.75'' (143\,au in the sky-plane) is WX\,Cha\,B, an M5 very low mass star (VLMS) of mass $0.18\pm0.06$\,M$_\odot$ \citep{daemgen2013}. WX\,Cha\,A is also a variable source at optical wavelengths, with variability in timescales from hours to years, and accretion luminosity with variability between 1.7\,L$_\odot$ and 3.5\,L$_\odot$ over a period of 2 months \citep{fiorellino2022}. Similarly to VW\,Cha\,A, the outer disk morphology remains unconstrained, although kinematic studies with CO lines in the near infrared have suggested that WX\,Cha\,A might be highly inclined \citep[87\,deg,][]{banzatti2015}.

\subsection{RW Aur}

RW\,Aur is a binary system in the Taurus star-forming region (SFR), located at 154\,pc \citep{gaia2021edr3}. The separation between A and B is 1.5'' (240\,au in the sky-plane). Each of the stars hosts its own circumstellar disk, first detected by \citet{Cabrit2006}, and later resolved by ALMA \citep{rodriguez2018, long2019}. Most recently, the morphology of the dust continuum emission was described in \citet{kurtovic2025a}, with dust disk sizes of 19\,au and 14\,au for A and B, respectively, with no annular structures detected in A, and a ring with a cavity of 4\,au in radius in B. The primary star is a K0 with a mass of 1.24\,M$_\odot$, while B is a K6.5 with a mass of 1.0\,M$_\odot$ (both with an estimated uncertainty of $0.05$\,M$_\odot$), with spectral types from \citet{rota2022} and stellar masses estimated from the CO outer disk rotation \citep{kurtovic2025a}. 

RW\,Aur was observed to be dynamically perturbed in \citet{Cabrit2006}, with an extended $^{12}$CO arc resulting from a close interaction \citep{Dai2015}. Later follow-ups with ALMA revealed additional arcs of gas emission detected in the CO $J=2-1$ transition, suggesting several past interactions \citep{rodriguez2018}. The most recent study of the orbital parameters suggests the last periastron was $295_{-227}^{+136}$\,yrs ago, triggering a bright and extended CO arc, truncating the disks, and tentatively exciting a warp in RW\,Aur\,A \citep{kurtovic2025a}. 

RW\,Aur\,A is also a highly active and episodic accretor, with pronounced dimming events at optical wavelengths, most likely associated to clouds of dust passing in front of the star along our line of sight \citep[see][]{rodriguez2013, chou2013, petrov2015, gunther2018, koutoulaki2019, lisse2022}, potentially associated to the disk \citep{facchini2016}. The high accretion around the primary star seems to be also connected to a large collimated outflow, detected as a jet in blueshifted and redshifted emission \citep[see][]{mundt1998, dougados2000, dougados2002, woitas2002, lopezmartin2003, beck2008, liu2012, takami2018, takami2020}. In comparison, RW\,Aur\,B is considerably less variable, although dimming events as large as 1.3\,mag in the V filter have been detected over a time span of days \citep{dodin2020}, also attributed to clouds of dust passing in front of the star. With an accretion rate of $\dot{M}\leq 5\cdot10^{-9}$\,M$_\odot$\,yr$^{-1}$, RW\,Aur\,B does not show high velocity forbidden line components, but a low velocity component in the optical forbidden line [O I] suggests the existence of a disk wind \citep{dodin2020}.

\subsection{DF Tau}

DF\,Tau is an equal mass binary system of mass 0.55\,M$_\odot$ and spectral type M2 \citep{allen2017}. These stars have an orbital period of 48\,yr, a moderate eccentricity of 0.2, and a semi-major axis of 97\,mas ($14$\,au when considering a distance of 140\,pc), derived by \citet{kutra2025}. Due to their small separation, the circumstellar disks have become highly compact, with sizes of $\leq3$\,au in the dust continuum emission from ALMA \citep{grant2024}. An analysis of the JWST/MIRI-MRS observations revealed that despite the very compact disk size, DF\,Tau\,A shows a bright spectrum dominated by H$_2$O lines, and a likely line-poor spectrum for DF\,Tau\,B. This relatively shorter separation binary, thus, provides a point of comparison for the moderate separation systems presented in this work.

\section{JWST-MIRI-MRS data reduction and analysis methods}\label{sec:obs:data}

\subsection{Observations and data reduction}\label{sec:obs:data:reduction}

RW\,Aur, VW\,Cha, and WX\,Cha were observed with the Mid-InfraRed Instrument \citep[MIRI,][]{rieke2015, wright2015, wright2023} with the Medium Resolution Spectrometer (MRS, \citealt{wells2015, argyriou2023}) mode on 24-July-2023, 7-August-2023, and 15-October-2023, respectively. These observations are part of the MINDS \citep[PID: 1282, PI: T. Henning,][]{kamp2023, henning2024}. A four-point dither was performed in the positive direction. The total exposure time was 27.2\,min per target, and target acquisition was not utilized in these observations. The observation details for DF Tau can be found in \citep{grant2024}. 
The MIRI-MRS data\footnote{The original data can be downloaded from the Mikulski Archive for Space Telescopes (MAST).} was reduced using a hybrid pipeline\footnote{The pipeline and associated documentation are available at \url{https://github.com/VChristiaens/MINDS}} \citep{minds_notebook}, combining routines from the standard JWST pipeline \citep{bushouse_1.13.4} using CRDS context 1224, and from the VIP package \citep{GomezGonzalez2017, Christiaens2023}. The pipeline is structured similarly to the standard JWST pipeline, but with a custom step to flag additional bad pixels.

The binaries in this work have separations between $0.7''-1.5''$, which allows them to be spatially resolved at the shortest wavelengths of MIRI/MRS. At longer wavelengths, the point-spread function (PSF) becomes wider, and the wings of the PSF of each source overlap with the position of the companions, which blends the flux of the sources. As one of the main goals of this work is to distinguish the spectra of primary and secondary, we apply a combination of forward modeling with a theoretical PSF and aperture photometry to disentangle the emission of each binary component. This is described in more detail in the Appendix \ref{sec:app:extract-spectra}. Before extracting the spectra, we calculated the separation between emitting sources by fitting their position at each wavelength in the band 1A. We allowed the center of each PSF to be a free parameter, and we calculated the center with the maximum likelihood with a Markov Chain Monte Carlo (MCMC) approach, similar to the procedure described in Appendix \ref{sec:app:extract-spectra}. After fitting the whole band 1A, we take the median separation in right ascension and declination to obtain the binary separation, which we fixed when fitting all the other bands from 1B to 4C.

\begin{figure*}[t]
  \centering
    \includegraphics[width=18cm]{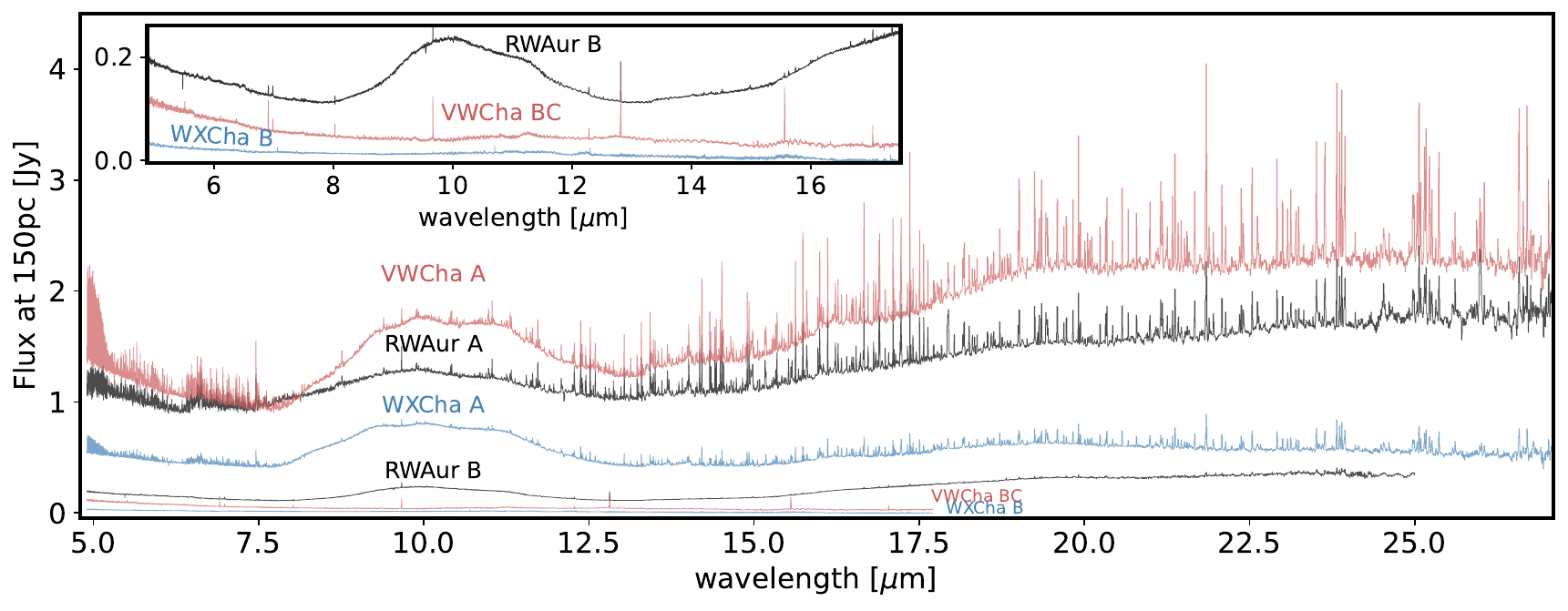}\\
    \vspace{-0.1cm}
    \caption{JWST/MIRI-MRS spectra for the three binary systems in our sample, scaled to 150\,pc for comparison. The spectra of the secondary sources are enlarged in the insert, showing the observed flux between the MIRI-MRS band 1A and 3C. }
    \label{fig:all_jwstmiri}
\end{figure*}

The isolated spectra of the primary and secondary disks are presented in Fig.~\ref{fig:all_jwstmiri}. In our forward modeled spectral extraction, the sensitivity in the secondary disks is limited by fringing in the detector, which prevented us from detecting faint molecules contributing to their spectra. This is further discussed in Sect.~\ref{sec:disc:limitations}. We estimate the continuum emission in the 1D spectra as in \citet{temmink2024}, which is the same approach that have been used in other studies \citep[e.g.,][]{grant2024, vlasblom2025}. In summary, the continuum is estimated iteratively with a Savitzky-Golay filter by fitting a third-order polynomial, masking spikes deviating by $2\,\sigma$ in the positive direction and $3\,\sigma$ in the negative direction of the spectra. The baseline of the filtered spectra is later determined using PyBaselines \citep{erb2022}, which we subtract from the observation to obtain the spectrum used to characterize the line emission.

\subsection{Modeling spectra with 0D slabs}\label{sec:obs:mcmc-slabs}

We fit the continuum-subtracted spectra with Local Thermodynamical Equilibrium (LTE) slab models in the range 13.4-25.4\,$\mu$m, to be comparable to previous works in the literature \citep[e.g.,][]{grant2023, grant2024, schwarz2024, temmink2024, gasman2025}. The spectroscopic data was obtained from the HITRAN database \citep{gordon2022}. Our slab calculation is the same as described in \citet{tabone2023}, which takes into account the line overlap for each species, which is needed when calculating the emission for lines that are near the optically thick regime \citep[e.g.,][]{grant2023, perotti2023, tabone2023}. We do not include line overlap between different molecular species. 
We fit the slab models to the spectrum of each disk using an MCMC approach with \texttt{emcee} \citep{emcee2013}, similar to that of \citet{grant2024} and \citet{perotti2025}. We searched for the combination of slabs that has the maximum likelihood of describing the continuum subtracted spectra. The considered wavelength range allows us to study the emission of C$_2$H$_2$, HCN, CO$_2$, OH, as well as multiple components of H$_2$O emission with a range of temperatures \citep[see][for additional discussions about the thermal structure of water in the MIRI-MRS wavelengths]{banzatti2023, temmink2024b, romeromirza2024, banzatti2025}. 
The MRS spectral resolution changes over its wavelength range, as several bands are stitched together to create the full spectra. When comparing the band 3B to 4C, there is a large difference in frequency resolution, ranging from about 2800 to 1600 \citep[see studies by][]{labiano2021, argyriou2023}. Therefore, we split our spectra by band, and we use a different uncertainties and spectral resolution for each. For the range covered by 3B-3C, we use an root-mean-square (rms) uncertainty of 2\,mJy, while an rms of 4\,mJy is used for bands 4A-4B, as in \citet{grant2024}. The spectral resolution for each band is taken from the most recent estimations \citep{pontoppidan2024, banzatti2025}.

The slab models for each molecule have 3 free parameters each: temperature ($T$), column density ($N$), and emitting area $\pi R_{slab}^2$ characterized by an emitting radius ($R_{slab}$). It is relevant to note that this area could have any shape or be at any location, as long as it conserves the area. The C$_2$H$_2$, HCN, and CO$_2$, are modeled with a single slab for each, and only in the region from 13.4\,$\mu$m to 17.7\,$\mu$m, as they do not provide detectable contributions at longer wavelengths. For H$_2$O and OH, we model them from 13.4\,$\mu$m to 25.4\,$\mu$m, and we do not include longer wavelengths as the S/N decreases dramatically. As in previous works \citep[e.g.,][]{grant2024, temmink2024b}, we model the water emission with three slab models, aiming to describe the emission from a hot ($\sim800$\,K), warm ($\sim400$\,K), and cold ($\sim200$\,K) temperature component, although the temperature is left as a free parameter. As linewidth, we use a $\Delta\text{V}=4.71\,\text{km}\,\text{s}^{-1}$ ($\sigma_\text{V}\approx2\,\text{km}\,\text{s}^{-1}$) as in \citet{Salyk2008, salyk2011a}, to remain comparable with the literature. We note that the effect of using constant or variable $\sigma_{\rm V}$ for fitting water emission with multiple slabs has been discussed in \citet{temmink2025}, with the main difference being a slight decrease in the values for column density when considering smaller $\sigma_{\rm V}$. The temperature of the slabs remains mostly constant within the uncertainties for different values of $\sigma_{\rm V}$ \citep[see Table D.4 in ][]{temmink2025}.

The slab models are calculated in the velocity rest-frame of the source, and then shifted using the source radial velocity, which is a free parameter of the model. We use the inverse of the parallax from Gaia DR3 \citep{gaia2021edr3} as the estimated distance to each system. The emission of all the slab models is added, and then binned to the frequency sampling of the MIRI-MRS using \texttt{spectres} \citep{carnall2017} before comparing it to the continuum-subtracted spectra. We optimize the minimum $\chi^2$ of the residuals as the likelihood distribution for our MCMC. We used a uniform prior for all the parameters. The minimum allowed temperature was 140\,K, while the maximum allowed radius was $10$\,au. The upper limit on column density was set to $10^{20}$\,cm$^{-2}$, except for the C$_2$H$_2$ and CO$_{2}$ of VW\,Cha and WX\,Cha, where an upper limit of $10^{18}$\,cm$^{-2}$ was used to prevent these molecules from trying to describe a pseudo-continuum. Thus, the results for them are lower limits. 
With the exception of these upper limits, which are further discussed in Sect.~\ref{sec:res:slabs}, the walkers did not interact with the boundaries. The MCMC ran until convergence was achieved for all parameters, which was typically obtained in the order of $4\cdot10^3$ steps for each walker, with $\times8$ the number of walkers as that of free parameters. After convergence, the MCMC was kept running for about $10^4$ steps for each walker, and only those steps were used to sample the posterior distributions.

\begin{figure*}[t]
  \centering
    \includegraphics[width=18cm]{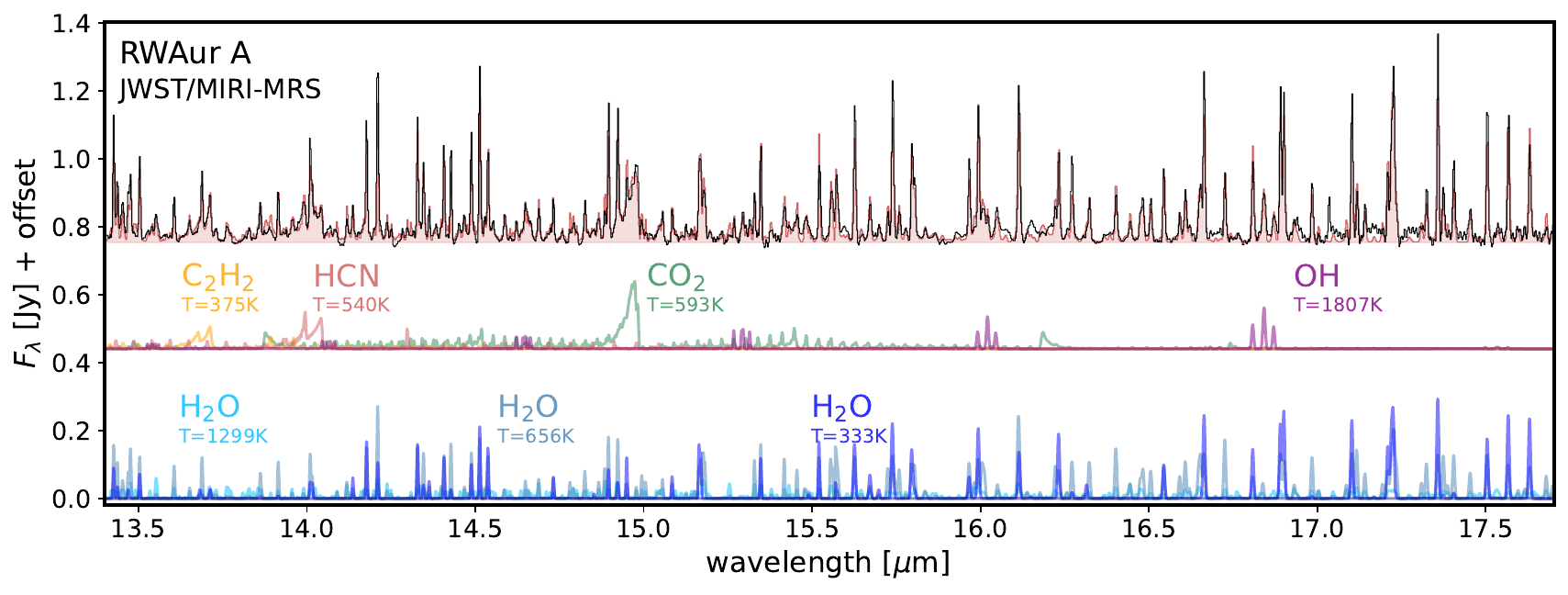}\\
    \includegraphics[width=18cm]{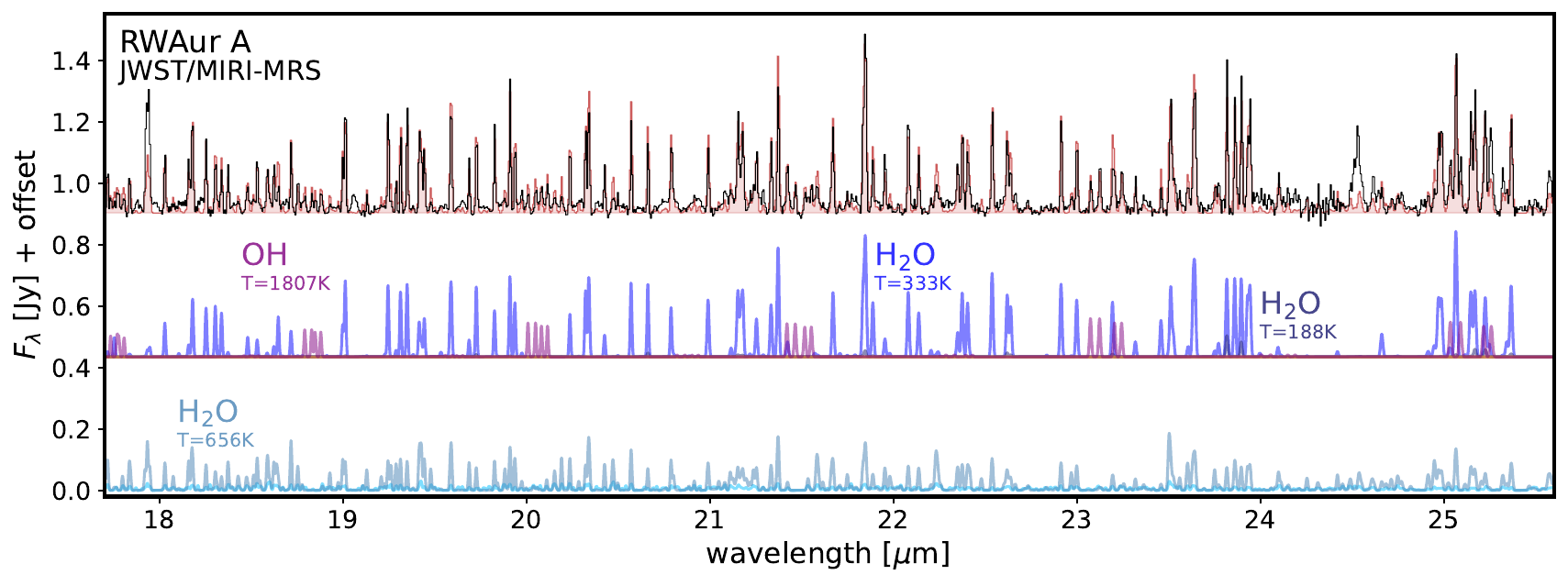}\\
    \vspace{-0.1cm}
    \caption{ The maximum likelihood model for each slab is shown in comparison to the continuum-subtracted spectrum of RW\,Aur\,A. These results are also presented in Tab~\ref{tab:mcmc-slabs}. }
    \label{fig:rwaur:spec-model}
\end{figure*}

\subsection{Extracting extended emission}\label{sec:obs:ext-emission}

We check for extended emission in all the channels containing H$_2$ emission, as well as the atomic lines that have been detected in other young stellar objects, such as those from the JOYS collaboration \citep{tychoniec2024}, including different transitions of ionized Iron (Fe), Nickel (Ni), and the noble gases Neon (Ne) and Argon (Ar).  
The lines detected in extended emission\footnote{The full list of lines where extended emission was checked is available in the ISO line list at \url{https://www.mpe.mpg.de/ir/ISO/linelists/}} are presented in Tab.~\ref{app:tab:lines_extended}. In this Section, we use the name ``channel images'' when referring to an image in one of the band cubes at a single wavelength, not to be confused with the MIRI-MRS Channels 1 to 4. We assume that every channel has contribution from a point source (the unresolved disk) and from extended emission (which might be zero or undetected in most channel images). We subtracted the point source emission by calculating a PSF for every channel image. This empirical PSF was calculated by taking the median of the 10 channel images with the nearest wavelengths to the line center, avoiding the four closest channels, as exemplified in Fig.~\ref{app:fig:median_psf}. This median PSF was rescaled to match the peak flux of each channel image, thus removing all the emission coming from a point source. This approach avoids including extended emission in the calculated PSF, as the considered channels are about 400\,km\,s$^{-1}$ apart. At the same time, the PSF is estimated from the nearest possible channels. For every potential extended emission line, we calculated a Moment 0 (integrated flux for every pixel) and a quasi-Moment 8 from fitting a quadratic function to the spectra of each pixel, using an adapted version of \texttt{bettermoments} \citep{bettermoments}.

\subsection{Limitations recovering the spectra of secondary disks}\label{sec:disc:limitations}

With the exception of DF\,Tau and VW\,Cha\,BC, the spatial resolution of the MIRI-MRS is enough to resolve the contribution to the mid-IR spectra from each disk. One of the challenges to studying the spectra of these objects is confidently recovering the emission of each disk without introducing PSF contamination from the nearby source. Despite having very water rich primary disks, no water is detected in the secondary disks in VW\,Cha and WX\,Cha, thus proving that the emission leak from one disk to the other is below the sensitivity of our recovered spectra. The limitations to our approach include the background subtraction, difficulty in avoiding problems such as the undersampling of the PSF, and fringing \citep{argyriou2023, law2023}. Due to these effects, the line detections in our spectra are limited to features brighter than 3mJy. This is a considerable limitation for studying the molecular emission of WX\,Cha\,B, which is a very low mass star. For this source, we are only able to detect emission corresponding to H$_2$, which is associated to extended emission around the WX\,Cha binary. Future analysis with forward spectra extraction in the detector-frame, instead of the cube-frame, could improve on effects such as the undersampling. For nearby binaries that fit within the same combined FOV from the four point dither ($<1.5$'' of separation), observing strategies that use target acquisition could be considered for future observations.

\section{JWST/MIRI-MRS Spectra} \label{sec:res:slabs}

\subsection{Overall spectra}

After separating the emission from primary and secondary disks, the most noticeable feature is the H$_2$O-dominated spectra of the primary stars, as presented in Fig.~\ref{fig:all_jwstmiri}. These disks are among the richest water line spectra of T-Tauri disks found to date. In contrast, all of the secondary disks are mostly line-poor, with continuum dominated spectra and a few spikes of flux corresponding mostly to extended emission. These findings are similar to those from the binary DF\,Tau, where most of the molecular emission is consistent with originating from DF\,Tau\,A \citep{grant2024}.

In each of the binary systems, the emission at the MIRI-MRS wavelengths is dominated by the disk around the primary sources, as the secondary disks are fainter in flux by a factor of more than 7 compared to the primary at all wavelengths. For VW\,Cha and WX\,Cha, where the separation of the disks is about $0.7''$, the spectra of the secondary disks could only be recovered up to band 3C ($17.7\,\mu$m). Starting from band 4A, only the primary disk flux was detected. In RW\,Aur, the larger separation between the binaries and the higher flux contribution from RW\,Aur\,B allowed us to distinguish the spectra of each disk up to band 4B. The spectra of the objects with resolved separations at the MIRI/MRS wavelengths are presented in Fig.~\ref{fig:all_jwstmiri}, where the disks around VW\,Cha\,B and C are shown in their combined contribution as VW\,Cha\,BC. Similarly, the spectroscopic binaries that compose VW\,Cha\,A are not spatially resolved. Since their flux might be coming from a single circumbinary disk, we analyze it as a single source.

\begin{figure*}[t]
  \centering
    \includegraphics[width=18cm]{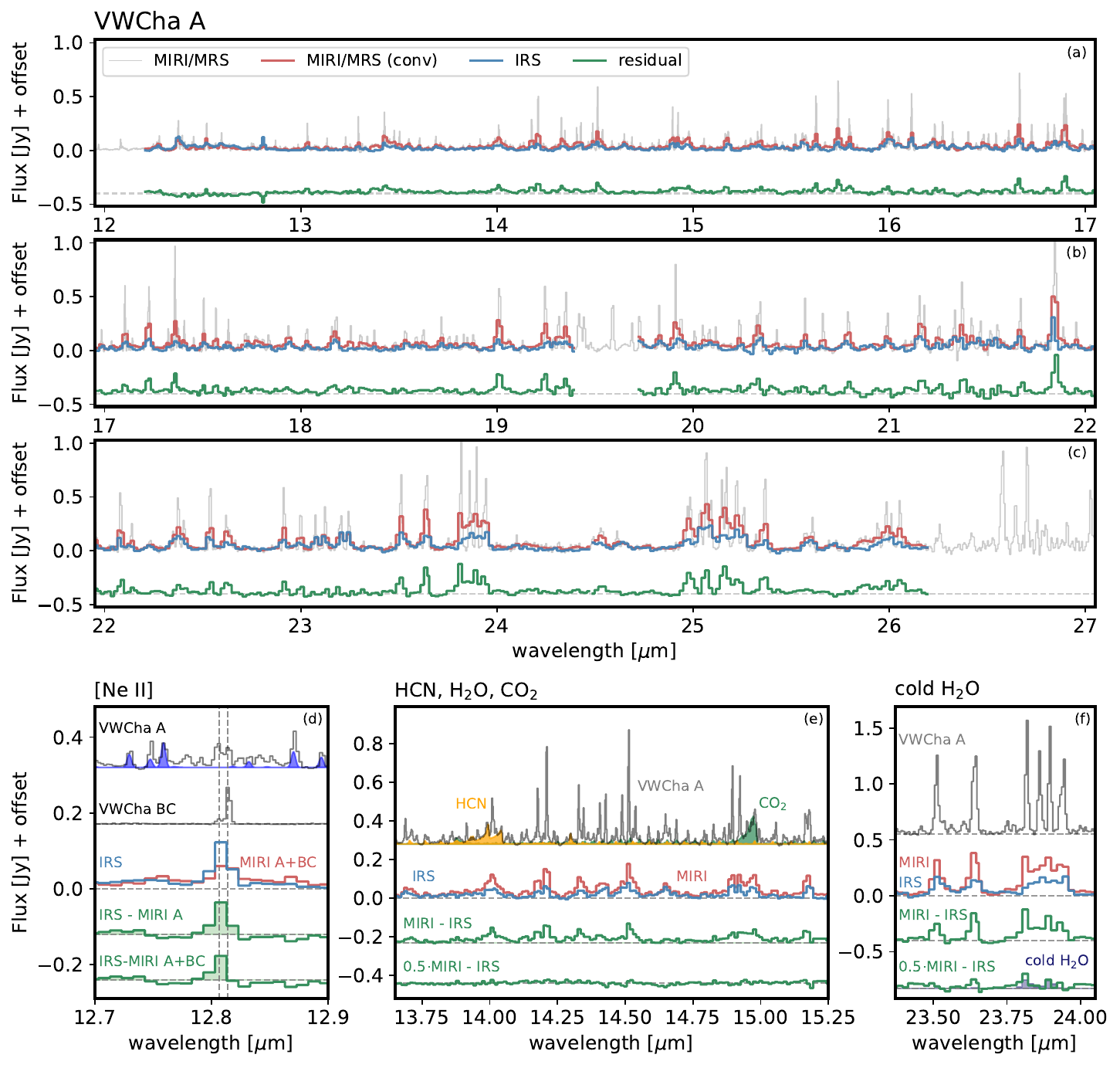}\\
    \vspace{-0.1cm}
    \caption{Comparison of the JWST/MIRI-MRS spectra and Spitzer/IRS in mode SH and LH. The panels \textbf{(a)}, \textbf{(b)}, and \textbf{(c)} show the comparison of MIRI-MRS with native resolution in grey, and in red after being convolved to the IRS resolution. The IRS spectra is shown in blue. The residuals from subtracting the IRS spectra to the MIRI-MRS are shown in green. Panel \textbf{(d)} is a zoom into the [Ne II] emission, with two vertical dashed lines showing the location of the [Ne II] at rest and the wavelength of [Ne II] with -180\,km\,s$^{-1}$ of blueshift. The emission of VW\,Cha\,BC is also shown, and convolved to the resolution of the IRS. Panel \textbf{(e)} is a zoom into the region with the main emission of HCN and CO$_2$. For comparison, the residuals from subtracting the IRS observation to the MIRI-MRS observation are shown both with their real flux, and with the MIRI-MRS flux scaled by 0.5. Panel \textbf{(f)} shows the region with the most prominent cold H$_2$O lines. Even after scaling MIRI-MRS by 0.5, there are still residuals consistent with a 200\,K slab.}
    \label{fig:vwcha:comp-spt-jwst}
\end{figure*}

In addition to the prominent water features at 7\,$\mu$m and above 12\,$\mu$m, the primary disks also show the characteristic silicate features at 10\,$\mu$m, and bright CO emission at 5\,$\mu$m. The silicate feature is also observed in RW\,Aur\,B, as well as faint emission that is consistent with rotational lines of H$_2$O and the $Q$-branch of CO$_2$, but its spectrum is otherwise dominated by fringing and undersampling \citep[see][Crouzet et al., subm.]{argyriou2023, law2023}, as shown in Fig.~\ref{fig:all_jwstmiri} and \ref{fig:app:rwaurb_manual_fit}. The companions VW\,Cha\,BC and WX\,Cha\,B have a similar problem, but differently from RW\,Aur\,B, no obvious molecular emission is identified, with only a continuum flux, extended emission, and ionized lines being confidently detected, as discussed in detail in Sect.~\ref{sec:res:noble_gas}.

\subsection{Slab models for molecular line emission}

We explored a combination of slab models that returned the maximum likelihood for describing the continuum subtracted spectra of the primary disks, as described in Sect.~\ref{sec:obs:mcmc-slabs}. The emission of HCN, OH, and CO$_2$ is well described with a single slab model in the three primary disks. For C$_2$H$_2$, RW\,Aur\,A and WX\,Cha\,A are described with a single slab, but in VW\,Cha\,A the model is consistent with having a negligible contribution from this molecule, with most of the flux at 13.7\,$\mu$m being contributed by different H$_2$O slabs. Thus, we do not report a confident C$_2$H$_2$ detection in this source. The maximum likelihood parameters for the slabs are presented in Tab.~\ref{tab:mcmc-slabs}. The maximum likelihood model for RW\,Aur\,A is in Fig.~\ref{fig:rwaur:spec-model}, while the models for VW\,Cha\,A and WX\,Cha\,A are in Fig.~\ref{fig:vwcha_specmodel} and \ref{fig:wxcha_specmodel} respectively. 

The only secondary disk with molecular emission is RW\,Aur\,B, which shows emission lines in the 15\,$\mu$m region consistent with H$_2$O, HCN and CO$_2$. No C$_2$H$_2$ emission is detected at the sensitivity of our spectra. Due to the low S/N of the detections in the original spectra, an MCMC approach to fit those lines is not feasible, and thus we only show a manually fitted model for comparison in Fig.~\ref{fig:app:rwaurb_manual_fit}. The shape of HCN and CO$_2$ is more consistent with low temperature slabs ($\sim300$\,K), in the case of HCN to match the triangular shape and low amplitude of the 13.99\,$\mu$m peak, and in the case of CO$_2$ to match the wavelength of the peak emission. However, these qualitative features do not replace a quantitative fit, and thus we only provide a comparison to these low temperature components as reference. For both molecules, a column density of $N=10^{17}$\,cm$^{-2}$ was used. 

An additional experiment was performed to reduce the effects of fringing and undersampling in the spectra of RW\,Aur\,B, and get a better visualization of its line emission in the 15\,$\mu$m region. These results are also presented in Fig.~\ref{fig:app:rwaurb_manual_fit}, and are described in detail in the Appendix \ref{sec:app:rwaurb_fit}.

\begin{figure*}[t]
  \centering
    \includegraphics[width=18cm]{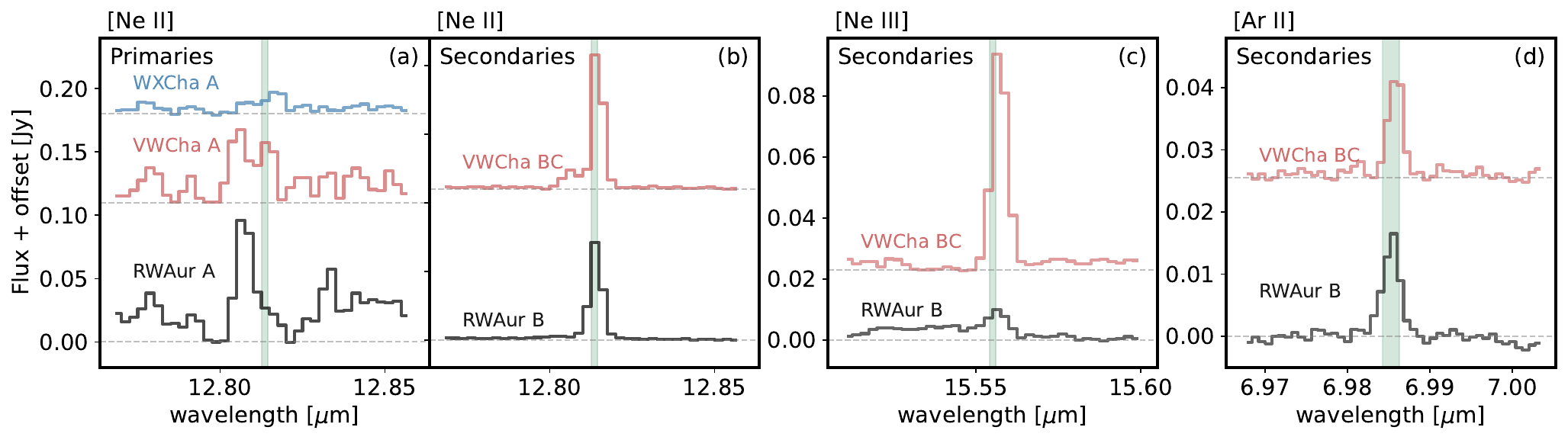}\\
    \vspace{-0.1cm}
    \caption{Detections of [Ne II] in all disks are shown in panels (a) and (b). Panel (c) and (d) show the emission of [Ne III] and [Ar II] in the secondary disks, respectively. A shaded region shows the rest wavelength of each ion transition. } 
    \label{fig:all-neon-argon}
\end{figure*}

\subsection{Thermal and density structure of H$_2$O emission}\label{sec:res:slabs:water}

Similarly to other sources with abundant H$_2$O lines \citep[e.g.,][]{gasman2023, xie2023, temmink2024, grant2024, romeromirza2024}, a gradient in the temperature of the H$_2$O slab models is needed to describe the observed emission. For VW\,Cha\,A and WX\,Cha\,A, the observed spectra is consistent with the three H$_2$O components, approximately representing a hot component ($\sim800$\,K), a warm ($\sim400$\,K), and a very extended colder component ($\sim200$\,K). The relation between temperature and radii of the three slabs is well described by a power law close to $R^{-0.5}$, similar to other disks in single-stellar systems \citep[e.g.,][]{temmink2024, romeromirza2024}. This power law relation has been demonstrated to be robust when using constant or variable values for $\sigma_{\rm V}$ \citep{temmink2025}. The column density of VW\,Cha\,A and WX\,Cha\,A also decreases as a function of emitting radius closely resembling a power law, similar to DF\,Tau \citep{grant2024}.

Interestingly, the model with three slabs of H$_2$O does not describe properly the spectrum of RW\,Aur\,A. When fitting with three slabs, we recover a temperature gradient as a function of emitting radii, but the column density remains mostly constant between the two slabs with higher temperature, similar to FZ\,Tau in \citet{romeromirza2024}. In this three slab model, the bulk of the H$_2$O flux is described by the moderate temperature H$_2$O slab, which considerably overestimates the H$_2$O line emission at 25\,$\mu$m (by more than 5$\sigma$ of difference between model and data). 
These results suggest that three slabs are not enough to describe the thermal structure of the H$_2$O emission, and thus, it is likely more structured than a single power law as a function of radius (see also work by Temmink et al., subm). 
We increased the complexity of the model with one additional H$_2$O slab, and we penalized the $\chi^2$ of the models that overestimated the emission at 25\,$\mu$m. These results are shown in Fig.~\ref{fig:rwaur:spec-model}. With 4 slabs, RW\,Aur\,A still has a more structured column density relation with the emitting radius than a single power law. If we considered the emitting radius to represent closely the physical radius where the molecules represented by the slab are located, then these results would suggest a tentative ring of increased column density at 1\,au, as detailed in Tab.~\ref{tab:mcmc-slabs}.

\subsection{Continuum and H$_\textit{2}$O variability}\label{sec:res:slabs:irs-miri}

RW\,Aur, VW\,Cha, and WX\,Cha were previously observed with Spitzer/IRS \citep{werner2004, houck2004} in its Short-High (SH) and Long-High (LH) mode \citep{lebouteiller2015}, and analyzed in \citet{carr2011}, \citet{pontoppidan2010b}, and \citet{salyk2011a}. Considering the faint continuum and molecular line emission from the companions in VW\,Cha and WX\,Cha (starting from 10\,$\mu$m, the flux ratio between A and B is larger than 25), it is safe to assume that all of the continuum emission in Spitzer/IRS was coming from the primary disks. When comparing the continuum of VW\,Cha and WX\,Cha between IRS and MIRI, we find their fluxes change less than 10\%, depending on the wavelength, despite being taken almost 15 years apart (as also presented in Fig.~\ref{fig:app:comp-spt-jwst}). The largest continuum variability is observed in RW\,Aur, where the continuum flux at 10\,$\mu$m decreased by about 50\% between IRS and MIRI-MRS. Continuum variability has been observed in other disks \citep[e.g.,][]{perotti2023, jang2024}, and in the case of RW\,Aur it seems to be physical in origin, and not explainable by combining the fluxes of RW\,Aur\,A and B together in the IRS observation. A near infrared flux variability has also been reported and analyzed by \citet{lisse2022} using a decade of observations of the IRTF/SpeX (0.7\,$\mu$m-4.1\,$\mu$m), and it is consistent with variability of a small ($R_{\text{em}}=0.04$\,au) and hot ($T=1650$\,K) accretion disk, potentially associated to an inner disk thermal instability \citep{lisse2022}.

The SH-LH mode of Spitzer/IRS had enough frequency resolution to start resolving H$_2$O lines \citep[see][for a detailed discussion]{pontoppidan2010b}, which we can compare to the emission observed by JWST/MIRI-MRS. 
We calculate and subtract the continuum of the IRS observations following the same procedure described in Sect.~\ref{sec:obs:data:reduction}, and convolve the MIRI-MRS spectra with a Gaussian kernel to reduce its frequency resolution. Further details about matching the two datasets are given in the Appendix \ref{sec:app:match-spt-jwst}. It is relevant to consider that the lower resolving power of Spitzer/IRS can also merge lines into a pseudo-continuum \citep[see Fig. 2 in ][]{jellison2024}. As exampled by RW\,Aur\,A and WX\,Cha\,A, this pseudo-continuum subtraction is not the dominant driver of differences between the spectra, but it represents a systematic source of uncertainty.

Despite having a similar continuum flux, the observation of VW\,Cha has a large line emission variability when comparing IRS to MIRI-MRS, as shown in the upper three panels of Fig.~\ref{fig:vwcha:comp-spt-jwst}, where the H$_2$O residuals are the most noticeable. After subtracting the IRS observation from MIRI-MRS, residuals at the location of the HCN and CO$_2$ at 14\,$\mu$m and 15\,$\mu$m are also observed, suggesting the variability exists across different molecular species. As a test, we scale the line emission of MIRI-MRS by a factor of 0.5, which is equivalent to reducing the emitting area of the molecules by half, but maintaining their temperature and column density. The results of this scaling are shown in panel (e) of Fig.~\ref{fig:vwcha:comp-spt-jwst} for the region between 13.6\,$\mu$m and 15.2\,$\mu$m, with a noticeable improvement in the amplitude of the residuals. At 24\,$\mu$m, in the region where the coldest H$_2$O (200\,K) starts contributing to the emission, a scale down by a factor of 0.5 is not enough to match the IRS observation, which suggest that the emitting area of cold H$_2$O was larger by more than a factor of $2$ during the MIRI-MRS observation.

The same comparison between IRS and MIRI-MRS was done for RW\,Aur and WX\,Cha. In both disks, the line emission does not show a detectable variability. These comparisons are presented in Figs.~\ref{fig:app:rwaur:comp-spt-jwst} and \ref{fig:app:wxcha:comp-spt-jwst}. For DF\,Tau, the only available observation with the IRS is in low spectral resolution mode \citep{lebouteiller2011}, which blends the water lines into the continuum, and prevents a comparison of emission lines as in the other binaries. Thus, we only show the comparison of continuum flux in Fig.~\ref{fig:app:comp-spt-jwst}.

\subsection{Ionized Neon and Argon}\label{sec:res:noble_gas}

Ionized Neon in the forbidden line [Ne II] at 12.81\,$\mu$m is detected in all the primary disks, as well as in the secondaries RW\,Aur\,B and VW\,Cha\,BC. In order to study the noble gas emission, including their total flux and velocity relative to the rest wavelength, we fit the spectra with two independent Gaussians, as described in Appendix \ref{sec:app:fit_noblegas}. These results are presented in Fig.~\ref{fig:vwcha:comp-spt-jwst} with a dashed line, as well as in Tab.~\ref{tab:noble_gas}.

In VW\,Cha\,A, we tentatively observe two peaks of [Ne II] emission. One of the emission lines is closely centered at the rest velocity of [Ne II], while the blueshifted component has a central velocity of $-175\pm3$\,km\,s$^{-1}$, also previously identified in \citet{pascucci2020}. 
In VW\,Cha\,BC, the [Ne II] emission has a double peaked profile in wavelength, with a blueshifted peak about 7 times fainter than the emission at rest velocity, whereas the two peaks have a similar amplitude in VW\,Cha\,A. 
For VW\,Cha\,BC, we find the blueshifted component is described by a Gaussian with a radial velocity of $-163\pm7$\,km\,s$^{-1}$, roughly consistent with the blueshifted component of VW\,Cha\,A at $-175\pm3$\,km\,s$^{-1}$, which was measured using the same approach. 
However, the blueshifted [Ne II] emission of VW\,Cha\,BC is considerably fainter than in VW\,Cha\,A, with a flux ratio of 7 between A and BC. In both cases, the flux was estimated from the Gaussian fit. Thus, the blueshifted emission detected in VW\,Cha\,BC could be attributed to the jet of VW\,Cha\,A, under the assumption that the jet axis is close to parallel to the position angle of VW\,Cha\,BC relative to A. 

Interestingly, a flux variability is detected in the [Ne II] emission of VW\,Cha when comparing IRS to MIRI, with the IRS flux being higher than that of MIRI by a factor of about 2.3 in total flux. This variability is not explained by spatially resolving VW\,Cha\,A from VW\,Cha\,BC, as combining their fluxes is not enough to match the flux observed with IRS. This variability is also presented in the lower left panel of Fig.~\ref{fig:vwcha:comp-spt-jwst}.

In WX\,Cha\,A, the [Ne II] shows a double peak emission consistent with blueshifted and redshifted jets, which are spatially resolved by MIRI-MRS (see Sect.~\ref{sec:ext:neon}). In RW\,Aur\,A, the blueshifted component dominates the [Ne II] emission, also consistent with the asymmetrical brightness of the extended jet.

In addition to [Ne II], the secondary disks VW\,Cha\,BC and RW\,Aur\,B also show bright [Ne III] and [Ar II] emission, as presented in Fig.~\ref{fig:all-neon-argon}. These lines are consistent with having a radial velocity close to the rest velocity, and thus they could be originating in an ionized wind component. 
The primary disks VW\,Cha\,A and RW\,Aur\,A do not show a detectable low velocity component in these ions, which supports that the emission is genuinely associated with the secondaries. 
When considering the flux ratio between the different ions, we find a [Ne II]/[Ar II] ratio of 4.0$\pm$0.5 and 2.5$\pm$0.4 for VW\,Cha\,BC and RW\,Aur\,B, respectively. For [Ne III]/[Ne II], we find ratios of 0.65$\pm$0.02 and 0.14$\pm$0.02. The lines [Ne III] and Ar[II] are not detected in any of our primary disks.

\subsection{Inner gas disk with VLT/CRIRES+}

All our primary disks are detected in CO emission at the shortest wavelengths of MIRI-MRS (4.92\,$\mu$m). In order to study the emission structure of this CO ro-vibrational emission, we use the high-spectral resolution observations in the M-band from VLT/CRIRES \citep{kaeufl2004} and VLT/CRIRES+ \citep{dorn2014, dorn2023}, which are able to recover the velocity structure of the CO in the inner disk \citep[e.g.,][]{brown2013, pontoppidan2011, bosman2019, banzatti2022, grant2024a}. These observations are detailed in Sect.~\ref{sec:app:criresp}. 

\begin{figure}[t]
  \centering
    \includegraphics[width=0.9\columnwidth]{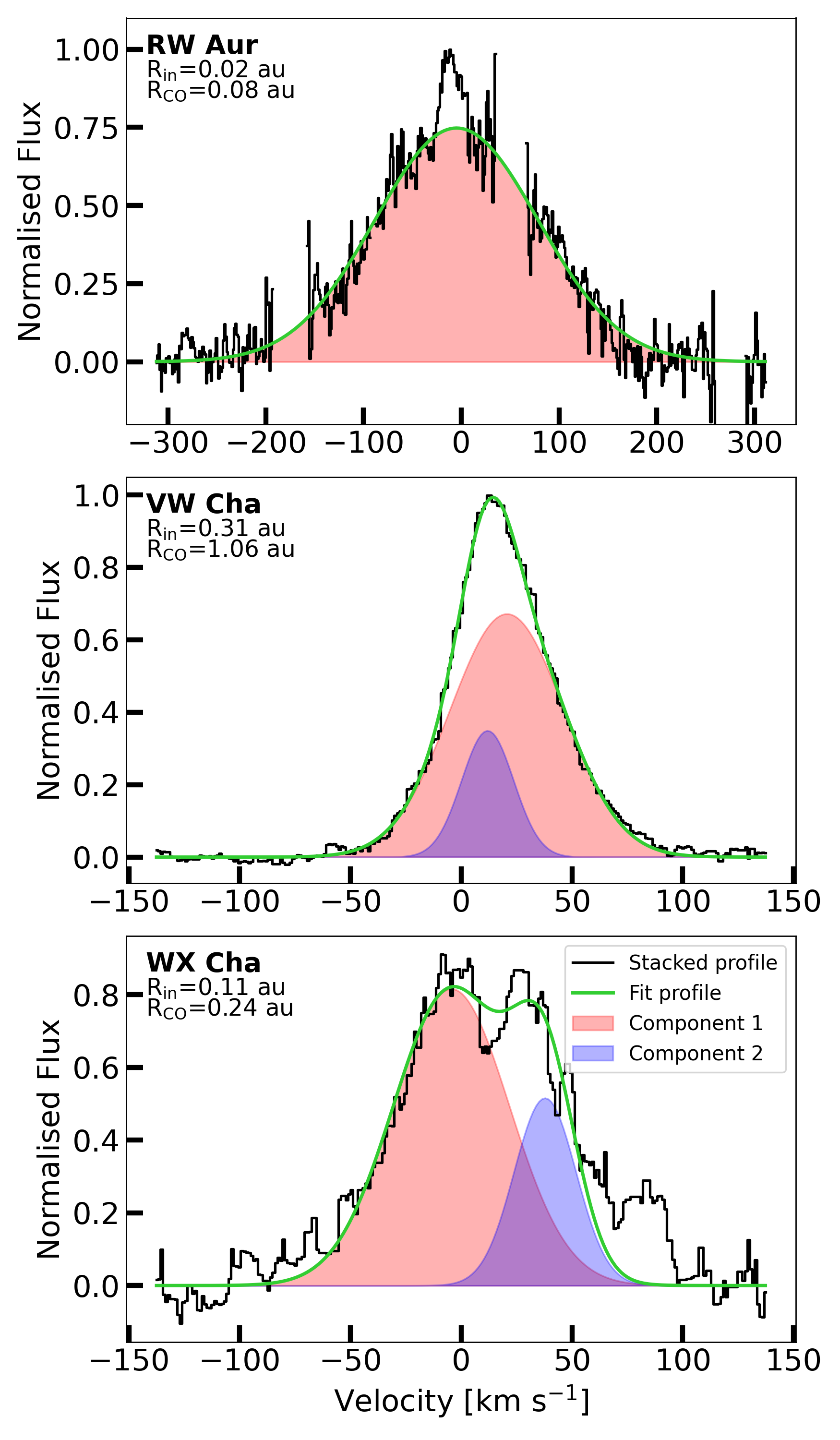}\\
    \vspace{-0.1cm}
    \caption{Line stacked CO profiles from CRIRES+ for VW\,Cha and WX\,Cha, and from CRIRES for RW\,Aur. }
    \label{fig:crires_co_profs}
\end{figure}

The frequency coverage in the VLT/CRIRES M-band ($\sim4.5\,\mu$m to $\sim5.0\,\mu$m) includes several lines from CO $\nu=1-0$ and $\nu=2-1$ transitions, which can be stacked to increase the S/N of their detection. The procedure for line stacking is described in Sect.~\ref{sec:app:criresp}, based on the work of \citet{temmink2024}, and the results are presented in Fig.~\ref{fig:crires_co_profs}. We fitted these lines using two Gaussian components, as these can describe broad-narrow components \citep{banzatti2022}, or double peaked Keplerian profiles. The fitting procedure is also described in Sect.~\ref{sec:app:criresp}. 

The profiles of both RW\,Aur\,A and VW\,Cha\,A are consistent with a broad Gaussian component, which is likely describing the Keplerian rotation of the inner disk. A narrow Gaussian component is also identified in VW\,Cha\,A, with a low velocity component that has been associated to a disk wind \citep{bast2011, brown2013, banzatti2022}. This emission is consistent with a previous finding of \citet{banzatti2023}, where tentative evidence for a blue-shifted wind was reported from the H$_2$O emission at 12.4\,$\mu$m. 
In WX Cha, the two Gaussian components describe a double peaked profile.

Using the combination of the fitted Gaussian components, we can begin constraining the characteristic emitting radii of the CO emission ($R_{CO}$), typically associated to the full width at half maximum (FWHM) of the velocity profile, as well as the inner radius of the gas disk ($R_{in}$), commonly calculated from the full width at $10\%$ of the emission (FW10). The transformation from velocity to a disk radius is done by using Kepler's laws \citep[see also][]{salyk2011b, banzatti2015, banzatti2022}, under the assumption of a fixed disk inclination and stellar mass. For RW\,Aur\,A, the inclination of the outer disk from ALMA is 55\,deg \citep{kurtovic2025a}, but in VW\,Cha and WX\,Cha these values are not as well constrained, due to the lack of high angular resolution millimeter observations. We use 60\,deg and 80\,deg for VW\,Cha and WX\,Cha, respectively, consistent with the results from \citet{banzatti2015}. The results for the Gaussian fit, $R_{CO}$, and $R_{in}$, are presented in Fig.~\ref{fig:crires_co_profs}. 

We find a good agreement between the values obtained from the CO emission in the M-band, and the hottest water component of each disk. For example, the characteristic emitting radius $R_{CO}$ of RW\,Aur is similar to that of $R_{\rm hot, H_2O}$ (see Tab.~\ref{tab:mcmc-slabs}), suggesting these two molecules are emitting from a similar region in the inner disk. In the case of VW\,Cha\,A, which is a circumbinary disk, the value for $R_{in}$ is the largest from the three primary disks, possibly due to a cavity opened by the spectroscopic binaries. However, this value should be revisited once the disk inclination is constrained by another instrument, such as ALMA. The large value for VW\,Cha $R_{CO}$ is also consistent with the large value for $R_{\rm hot, H_2O}$ and $R_{\rm warm, H_2O}$ found with the slabs.

\begin{figure*}[t]
  \centering
    \includegraphics[width=17cm]{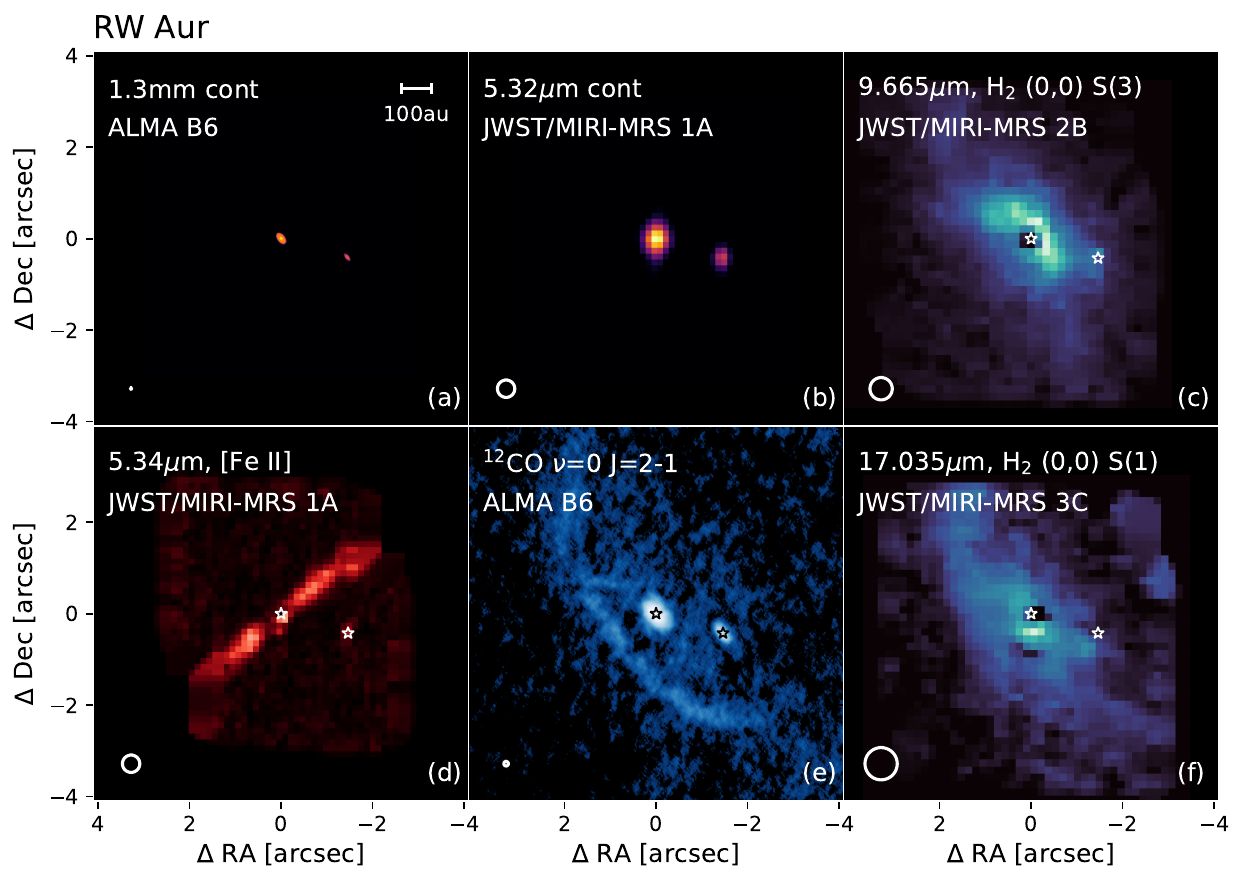}\\
    \vspace{-0.1cm}
    \caption{The $\approx$1000\,au environment of RW\,Aur at different wavelengths, as observed by ALMA and JWST. Panel (a): compact dust continuum emission at 1.3\,mm wavelength \citep[see][for a detailed study]{kurtovic2025a}. Panel (b): continuum emission from MIRI-MRS subband 1A. Panel (c): Peak emission map of H$_2$ S(3). Panel (d): Peak emission map of [Fe II]. Panel (e): Moment 0 map of the $^{12}$CO emission at 230.538\,GHz, as observed with ALMA. Panel (f): Peak emission map of H$_2$ S(3). }
    \label{fig:rwaur:gallery_v2}
\end{figure*}

\section{Extended emission} \label{sec:ext:extended}

After subtracting the flux contribution of a point source, as described in Sect.~\ref{sec:obs:ext-emission}, we detect additional emission in several H$_2$ and ionized atoms transitions. In the case of RW\,Aur and DF\,Tau, where high angular resolution observations at millimeter wavelengths are available \citep{grant2024, kurtovic2025a}, the interpretation of the extended emission can be done by comparing with the colder dust and gas traced by ALMA. We compare these two sources in Fig.~\ref{fig:rwaur:gallery_v2} and Fig.~\ref{fig:app:dftau-alma-jwst}. For VW\,Cha and WX\,Cha, however, the interpretation is limited to the observations from JWST/MIRI, which we describe in this Section.

\subsection{Extended Ionized Neon}\label{sec:ext:neon}

\begin{figure*}[t]
  \centering
    \includegraphics[width=\linewidth]{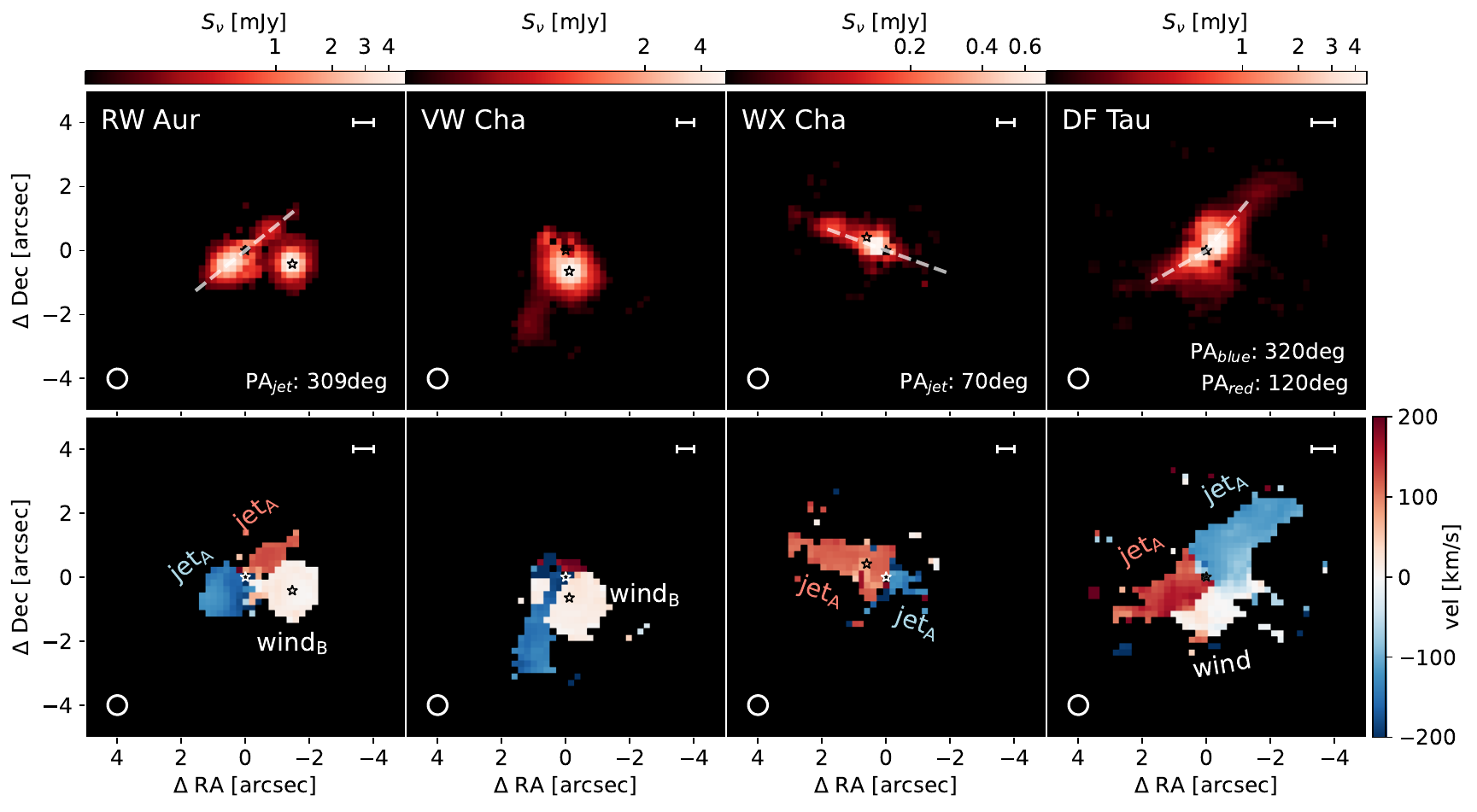}\\
  \vspace{-0.1cm}
  \caption{Peak brightness and velocity at peak for the [Ne II] emission at 12.814\,$\mu$m. A dashed line shows the estimated position angle of the jet emission, which was only manually fitted for WX\,Cha and DF\,Tau. }
  \label{fig:all-neon-jets}
\end{figure*}

Among the atomic lines that are detected in extended emission, the ionized Neon [Ne II] in its 12.814\,$\mu$m transition is spatially resolved in a jet-like structure in RW\,Aur, WX\,Cha, and DF\,Tau. In RW\,Aur and WX\,Cha, the high-velocity component is being launched from the primary disk (see Fig.~\ref{fig:all-neon-jets}), while in DF\,Tau we do not have enough angular resolution to resolve which of the disks is launching the jet, although it is most likely originating from DF\,Tau\,A (see Sect.~\ref{sec:disc:noble-gas}). 
In contrast, the [Ne II] emission is spatially compact around the companions RW\,Aur\,B and VW\,Cha\,BC, and closely centered at the expected rest frequency of the systems, thus suggesting it might be related to a low-velocity component, possibly tracing radiative or wind-like origin. Such different behavior in outflow emission has also been observed in younger binaries \citep{tychoniec2024}. 

An extended emission structure is detected in VW\,Cha in blueshifted emission with a position angle close to 160\,deg. However, this emission is not at the same velocity as that measured in Sect.~\ref{sec:res:noble_gas}. 
As no redshifted counterpart is detected, confirming this blueshifted emission as originating from a jet is more challenging than in the other systems. Furthermore, \citet{bally2006} suggested an outflow associated to VW\,Cha with a position angle of 90\,deg, which does not coincide with the orientation observed by MIRI. Thus, we present this detection as tentative.

In all our binary systems, the [Ne II] jet is asymmetric in brightness when comparing the blueshifted to the redshifted component. In WX\,Cha, the jet position angle (PA$\approx$70\,deg to the redshifted side) is similar to that of the companion (PA$\approx 52$\,deg), and it is also brighter in this direction. In RW\,Aur, the jet is detected in several tracers including Iron, Nickel, Neon and Argon, and it will be further analyzed in a future work.

The most extended [Ne II] jet is observed in DF\,Tau, with redshifted and blueshifted components extending further than 2'' from the disks ($\approx 300$\,au). This jet had been previously reported in the optical \citep{uvarova2020} and NIR \citep{dodin2025}. Each side of the jet has a different position angle (PA$_{\text{red}}-\text{PA}_{\text{blue}}\approx20$\,deg), consistent with the findings of \citet{dodin2025}. We detect a tentative change of position angle along the jet extension, following a clock-wise deviation as a function of increasing distance from the system. When observed with ALMA in Band 6, DF\,Tau shows emission in the $^{12}$CO $\nu$=0 J=2-1 emission with a blueshifted and redshifted components, most likely originated from the primary disk rotation (see Fig.~\ref{fig:app:dftau-alma-jwst}). When comparing the position angle of the $^{12}$CO around DF\,Tau\,A to that of the [Ne II] jet, we confirm they are neither parallel (thus, $^{12}$CO of ALMA is not tracing an outflow), nor perpendicular, suggesting a misaligned jet launching axis relative to the outer Keplerian disk of DF\,Tau\,A.

\subsection{Extended $H_2$ emission}\label{sec:disc:h2}

Extended emission is detected in RW\,Aur and VW\,Cha for all the H$_2$ transitions covered by the MIRI/MRS, spanning from H$_2$\,S(1) to H$_2$\,S(8). In RW\,Aur, the transition H$_2$ S(1) shows an arc-like morphology, which changes at the higher energy transitions of H$_2$ into a bowl-like shape (see Fig.~\ref{fig:rwaur:gallery_v2} and \ref{fig:rwaur_extended}). In VW\,Cha, the emission has a consistent bowl-like morphology for every H$_2$ transition, as shown in Fig.~\ref{fig:vwcha_extended}. WX\,Cha is only confidently detected in extended emission in the low energy transitions of H$_2$, from S(1) to S(3), with tentative detections of emission in S(4) and S(5), as shown in Fig.~\ref{fig:wxcha_extended}. For DF\,Tau, we do not detect extended H$_{2}$ emission to the sensitivity and angular resolution of MIRI. When compared to the [Ne II] extended emission, we find the position angle of the H$_2$ bowl and [Ne II] jet are close to perpendicular for RW\,Aur and WX\,Cha.

The transitions from H$_2$\,S(5) to H$_2$\,S(8) are all within the wavelength range of the subbands 1A to 1C, which results in Moment 0 images with the same pixel size, and comparable angular resolution. Under the assumption that the column of H$_2$ contributing to the emission between S(5) to S(8) has a single temperature and column density within the area covered by each pixel, we can fit these higher energy transitions of H$_2$ in each of them with a slab model. 
The details about this fitting procedure and data treatment are explained in Appendix \ref{sec:app:fit_h2}. In pixels with high uncertainty and low flux contribution from H$_2$, the values for $T_{\text{H}_2}$ and $N_{\text{H}_2}$ can become unrealistic and deviate largely from the neighboring pixels, as the solution is not constrained at all in the absence of flux. Thus, we apply a filter by excluding solutions with temperatures lower than 500\,K or higher than 2000\,K, as well as column densities below $10^{17}$ or higher than $10^{19}$ in units of cm$^{-2}$. The fitted images and results are shown in Fig.~\ref{fig:h2model_highS}, for RW\,Aur and VW\,Cha, the two disks detected in H$_2$ emission between S(5) to S(8).

We find higher temperatures in the map of RW\,Aur than in that of VW\,Cha, with a median temperature of 1300\,K in RW\,Aur, and 800\,K in VW\,Cha. In RW\,Aur, the temperatures are consistently higher along the axis parallel to that of the jet, potentially tracing the jet cavity \citep[e.g.,][]{delabrosse2024, tychoniec2024, pascucci2025}. 
Considering the recovered value for column density $N_{\text{H}_2}$ and the size of the emitting area are known, we can estimate a map for the number of molecules per pixel, which can then be used to calculate the amount of mass. By integrating the mass of all pixels in the filtered images, we estimate the total mass of H$_2$ emitting in the transitions S(5) to S(8) is $9.2\cdot10^{-8}\,$M$_\odot$ in the RW\,Aur system, and $8.3\cdot10^{-8}\,$M$_\odot$ in VW\,Cha. It should be noted that this mass corresponds to the hot emitting H$_2$ ($>500\,$K), as we are only fitting the higher energy transitions, while most of disk and environment mass is contained in in a colder H$_2$ \citep[e.g.,][]{pascucci2013}. Obtaining meaningful uncertainties from these models values is challenging, as the pixel flux uncertainty is variable with pixel position over the image and also with the specific molecular transition.

\section{Discussion} \label{sec:disc}

\subsection{Considerations about the origin of binary systems}

Binaries are typically thought to have been formed together from the same cloud core or filament, thus having the same initial chemical composition. However, this is not the case for binaries that have recently become bound through a stellar capture, in which case the chemical composition of the inner disk could be different due to different ages or initial compositions. A third scenario should also be considered, which is when the interacting stars are not gravitationally bound, but rather in a high eccentricity orbit (also known as hyperbolic fly-bys). In such a single interaction event, simulations have demonstrated that companions can capture material from the circumstellar disk around the primary star for certain periastron distances \citep[e.g., ][]{Dai2015, Cuello2019}. These three scenarios, formed together, recently bound, or fly-by, are not necessarily mutually exclusive of each other, and all of them could have an impact in the inner disk molecular emission that we observe in the binary systems.

\subsection{The H$_2$O emission of binary systems}

Previous studies have suggested a connection between water rich sources and dust radial drift, as the pebbles bring ices from the outer disk and replenish the inner disk with H$_2$O \citep[see][]{banzatti2020, banzatti2023b}. The interaction with external companions can increase the rate of radial drift \citep{zagaria2023}, which should be happening to primary and secondary disks in our systems. However, the lack of prominent H$_2$O lines in the secondary disks suggests that additional mechanisms might be playing an important role in shaping their inner disk mid-IR signatures.

\subsubsection{RW\,Aur}
\label{sec:disc:rwaur}

One of the most interesting comparisons of the detected molecular emission comes from the RW\,Aur system, where the stars have similar stellar mass and outer disk size, but show completely different inner disk spectral features. RW\,Aur\,A has an H$_2$O-rich spectrum, with evidence of structured column density distribution, while RW\,Aur\,B only shows weak emission lines from HCN, CO$_2$, and H$_2$O, all of them consistent with a low temperature component ($\approx$300\,K). This difference might be connected to the morphology of the disks, as revealed by ALMA at high angular resolution. The disk of RW\,Aur\,A is consistent with being a full disk (with a spatial resolution of 4\,au, cavities larger than 2\,au would have been detected), with signs of having a warp in its inner 3\,au. RW\,Aur\,B, on the other hand, shows a 4\,au cavity in dust continuum emission \citep{kurtovic2025a}. A cavity of a few astronomical units in radius can change the flux ratio between CO$_2$ and H$_2$O, with an increasingly brighter CO$_2$ Q-branch compared to the H$_2$O lines for larger cavity sizes \citep{vlasblom2024}. Thus, the 4\,au cavity detected with ALMA is consistent with the weak emission lines detected in RW\,Aur\,B (see Fig.~\ref{fig:app:rwaurb_manual_fit}), at least qualitatively. Further dedicated thermochemical modeling could explore if the 4\,au cavity is also enough to explain the low amplitude of these lines in comparison to RW\,Aur\,A, or if additional mechanisms are needed to explain this difference.

When it comes to the origin of the binary system, RW\,Aur A and B show evidence of being in a bound orbit, but the scenarios or stellar capture or material capture have not been excluded \citep[see ][]{kurtovic2025a}, with the extended $^{12}$CO emission arcs being evidence of several past close interactions. Thus, it remains a possibility that the disks have an underlying different composition, but this hypothesis by itself is not enough to explain the difference in the temperature of the detected molecules.

\begin{figure*}[t]
  \centering
    \includegraphics[width=15cm]{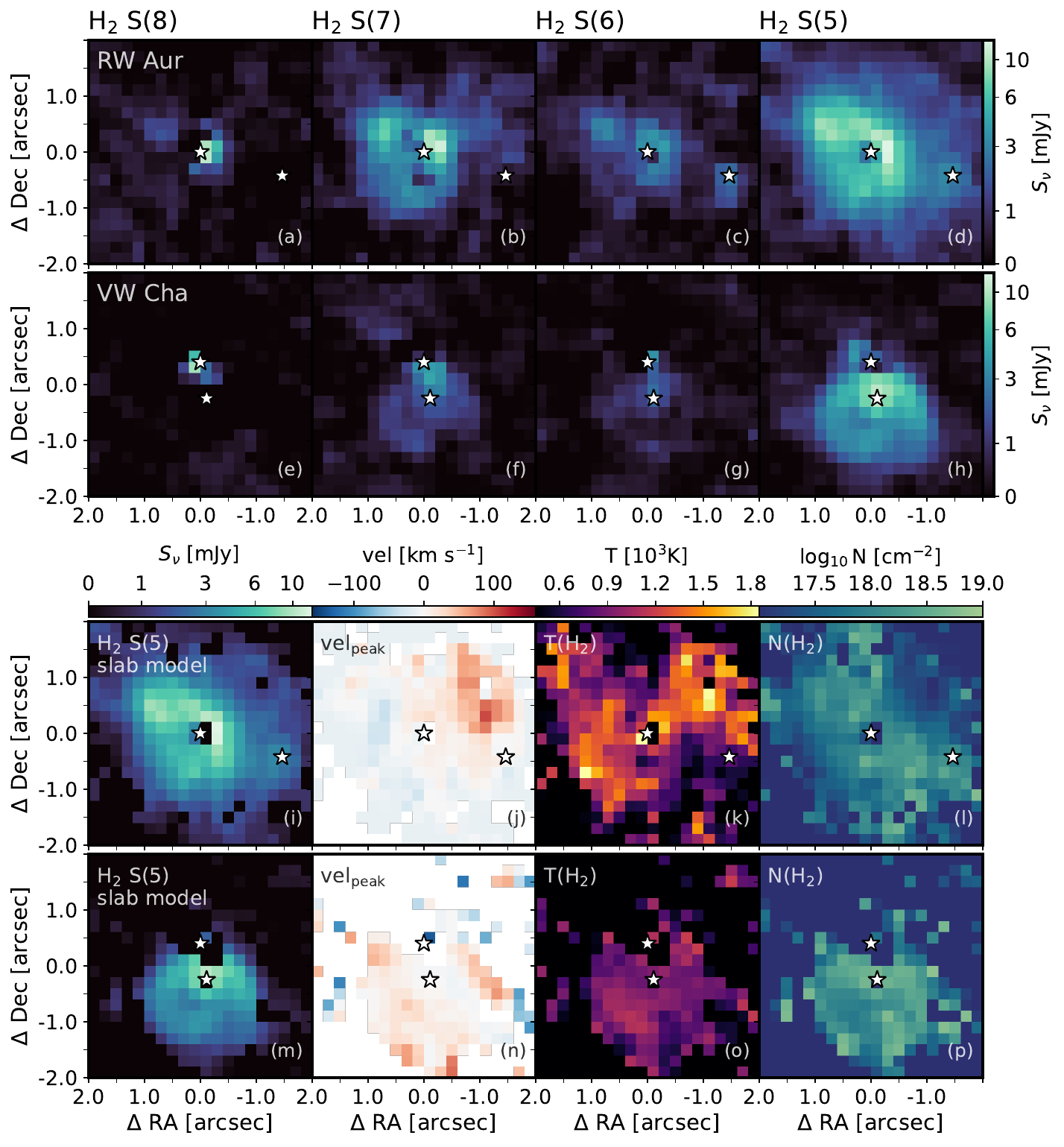}\\
    \vspace{-0.1cm}
    \caption{Pixel-by-pixel model for the H$_2$ in the transitions S(5) to S(8) toward RW\,Aur (emission in panels a-d) and VW\,Cha (emission in panels e-h), all of them in the MIRI-MRS field of view between band 1A to 1C. The upper rows shows the resampled H$_2$ images (panels a-h), while the lower rows show the best model results (panels i-l for RW\,Aur, and panels m-p for VW\,Cha). Panels i and m show the best model for H$_2$ S(5), panels j and n show the centroid velocity in the same line, panels k and o show the maximum likelihood slab temperature, and panels l and p show the maximum likelihood slab column density. }
    \label{fig:h2model_highS}
\end{figure*}

\subsubsection{VW\,Cha and WX\,Cha}

For the Chameleon sources, VW\,Cha and WX\,Cha, no high angular resolution ALMA observations are available to constrain the properties of the outer disks. Thus, establishing relations between inner disk mid-infrared spectra and outer disk properties remains speculative. In the following, we outline potential scenarios to explain the differences in spectral features between primaries and secondaries in both systems. 

In VW\,Cha, the emission from the secondary VW\,Cha\,BC is coming from two stars, which are indistinguishable to the angular resolution of MIRI-MRS, complicating the interpretation even further. Considering the derived masses from \citet{daemgen2013} for the dominant component of A and BC, we find a mass ratio of $q\approx1.3$, similar to the mass ratio of of RW\,Aur, and once again raising the question about the difference in H$_2$O luminosity. 
One possible explanation could be the proximity of VW\,Cha\,B and C as the responsible for highly truncating the disks and depleting them of almost all their material. However, the separation of VW\,Cha\,B-C is similar to that of DF\,Tau\,A-B, with also similar spectral types and stellar mass (DF\,Tau\,A and B are M2 and 0.55\,M$_\odot$), and DF\,Tau\,A shows a bright forest of H$_2$O lines in its MIRI-MRS spectra \citep{grant2024}. Further understanding of the difference in the H$_2$O emission will require information about the outer disk of each star, as well as the orbits in this quadruple system (A is a spectroscopic binary, and with B-C there are four stars), all of them recoverable with ALMA at high angular resolution. 
Differently from RW\,Aur, the extended emission of VW\,Cha shows no evidence of recent dynamical interactions in the extended emission of MIRI-MRS, which contributes to exclude scenarios such as recent stellar capture or unbound interactions. However, VW\,Cha is missing dedicated spatially resolved observations of cold gas tracers, such as $^{12}$CO at millimeter wavelengths, which could provide constraints for possible dynamically perturbed material.

In the WX\,Cha system, the difference between the H$_2$O emission between primary and secondary is less surprising, as WX\,Cha\,B is a very low-mass star of stellar type M5 and mass of 0.18\,M$_\odot$ \citep{daemgen2013}. These very low-mass objects commonly show very faint H$_2$O in comparison with the molecular emission coming from carbon species \citep[e.g.,][]{tabone2023, arabhavi2024, arabhavi2025b, arabhavi2025a, kanwar2024}. Due to the limitations of our procedure to extract the spectra, those molecular lines are not detected to our sensitivity (about 4\,mJy for the secondaries). Thus, WX\,Cha\,B could have a carbon-rich spectrum, but none of those lines are brighter than 4\,mJy at peak. For comparison, if WX\,Cha\,B had the same molecular emission of J1605 \citep{tabone2023}, we would have at least detected its C$_2$H$_2$.

\subsection{Ionized Neon and Argon in the binaries}\label{sec:disc:noble-gas}

A high velocity component of the [Ne II] emission is detected in all the primary disks, consistent with a jet origin (see Fig.~\ref{fig:all-neon-argon} and Fig.~\ref{fig:all-neon-jets}), as it has also been observed in other sources with MIRI \citep[e.g.,]{delabrosse2024, narang2024, tychoniec2024, pascucci2025}. In binary systems, these jets can reveal information about past disk-companion interactions. For example, in RW\,Aur\,A, the jet is misaligned relative to the 1.3\,mm continuum disk by at least 10\,deg, supporting a scenario in which the outer disk was warped relative to the inner disk during the last interaction with RW\,Aur\,B about 300\,yrs ago \citep{kurtovic2025a}.

In DF\,Tau, the binary separation is only 13$\pm$1\,au (about 94\,mas in the sky), with an orbital period of 46.1$\pm$1.9\,yrs \citep{allen2017}. Thus, the gravitational perturbations of DF\,Tau\,B are a likely explanation for the misaligned blue and redshifted [Ne II] jet components. 
The jet of DF\,Tau has been studied before with observations of the Hubble Space Telescope \citep[HST, ][]{uvarova2020}, but the proper motion has not yet been measured. We can estimate the jet velocity under certain assumptions. 
The disk of DF\,Tau\,A is consistent with being highly inclined from the ALMA $^{12}$CO $J=2-1$ observation (see Fig.~\ref{fig:app:dftau-alma-jwst}), with a possible inclination of 67\,deg to explain the Keplerian rotation of the ro-vibrational $^{12}$CO at 5\,$\mu$m \citep{grant2024}. Considering the line-of-sight velocity of the jet is about 100\,km\,s$^{-1}$, if the jet is being launched close to perpendicular to the disk, then the inclination of the jet would be about 23\,deg, and the proper motion velocity would be close to 0.35''\,yr$^{-1}$. Alternatively, if we assume as a guess that the launch velocity of the jet is 200\,km\,s$^{-1}$, then the proper motion would be about 0.25''\,yr$^{-1}$. Under the additional assumption that the launch velocity has been constant over the last binary orbit, then the full wiggle of the jet would have a period of 16.1'' or 11.5'', respectively for the both assumed scenarios. These distances are much larger than the field of view of MIRI-MRS, and dedicated observations would be needed to demonstrate if DF\,Tau\,B is indeed the cause of the wiggling. For the other binary systems, the larger binary separation between primary and secondary should not produce observable features related to dynamical perturbations. This is the case of RW\,Aur and WX\,Cha, where blueshifted and redshifted components are consistent with being parallel to each other to the angular resolution of MIRI-MRS.

In the secondary disks VW\,Cha\,BC and RW\,Aur\,B, the emission of Neon and Argon emission is at low velocity, and consistent with having unresolved velocity linewidths at the spectral resolution of MIRI-MRS \citep[resolution of 90$\sim$100\,km\,s$^{-1}$ for these transitions][]{banzatti2023}. The [Ne II] emission has been detected in tens of sources with Spitzer/IRS surveys \citep{pascucci2007,lahuis2007,najita2010, guedel2010, espaillat2013}, and later studies at high spectral resolution allowed to distinguish between wind and jet origins \citep{herczeg2007, vanboekel2009, pascucci2009, sacco2012, pascucci2020}. However, the detection of [Ne III] remained rare with Spitzer \citep{lahuis2007, najita2010, szulagyi2012, espaillat2013}. From an observational perspective, the emitting wavelength of [Ne III] at 15.555\,$\mu$m overlaps with a few H$_2$O lines from a hot component (600\,K), which makes it challenging to isolate in H$_2$O-rich sources, such as the primary disks in our binary systems. The line-poor disks allow for an easier detection, with a handful of Class II sources having confirmed detections prior to JWST, such as Sz\,102 \citep{lahuis2007}, TW\,Hya \citep{najita2010}, CS\,Cha, SZ\,Cha and T\,54 \citep{espaillat2013}, and now MIRI-MRS has expanded this sample to include T\,Cha \citep{bajaj2024}, SY\,Cha \citep{schwarz2025} and MY Lup \citep{salyk2025}. To this group of sources, we now include VW\,Cha\,BC and RW\,Aur\,B.

The detection of [Ne III] is relevant to study the high-energy radiation environment of the inner disk, as ionizing [Ne II] and [Ne III] requires extreme ultraviolet (EUV) or X-rays. 
In the case of a wind, the flux ratio between [Ne III] and [Ne II] has been suggested to distinguish between EUV dominated and X-ray dominated high-energy radiation \citep{hollenbach2009, szulagyi2012, sellek2024}. A flux ratio between these lines closer to $F_{\text{[Ne III]}}$/$F_{\text{[Ne II]}}\sim1$ favors a scenario where a strong EUV field is ionizing the Neon, while $F_{\text{[Ne III]}}$/$F_{\text{[Ne II]}}\sim0.1$ are more consistent with ionization produced by X-rays \citep{hollenbach2009, ercolano2010}. Disks have so far generally showed values inconsistent with a ratio of 1 \citep{najita2010, szulagyi2012, espaillat2013, bajaj2024, espaillat2024}, suggesting that X-ray ionized winds are more frequent. An exception is found in SZ\,Cha \citep{espaillat2013}, which firstly showed a $F_{\text{[Ne III]}}$/$F_{\text{[Ne II]}}\sim1.0$ with Spitzer/IRS, and a later follow-up with MIRI-MRS revealed significant variability in the line fluxes, with their ratio now consistent with X-ray ionization \citep{espaillat2023}. 
Within this context, RW\,Aur\,B, with a flux ratio of $0.14\pm0.02$, looks like a typical case of a X-ray ionized wind. 
On the other hand, VW\,Cha\,BC, with a ratio of $0.65\pm0.02$ would suggest EUV ionization similar to previously observed for SZ\,Cha in \citet{espaillat2013}.

The presence of strong ionized noble gas emission around the secondaries may have a common origin in relation to their weak molecular line emission. \citet{pascucci2020} showed that for sources with a low-velocity [Ne II] component, the line luminosity grew with increasing mid-infrared spectral index, often indicative of an increasingly cleared inner disk. This is consistent with our interpretation of a small 4\,au cavity driving the weak molecular emission in RW\,Aur\,B (see Sect.~\ref{sec:disc:rwaur}), and a larger sample of sources could test this hypothesis.

Finally, the ratio of [Ar II] and [Ne II] has also been used to explore the ionizing spectrum \citep{szulagyi2012, bajaj2024}. \citet{hollenbach2009} suggested that the [Ne II]/[Ar II] ratio should be $\sim1$ in gas ionized by EUV or soft X-rays, rising to $2.5$ in gas ionized by hard X-rays. \citet{sellek2024} showed the latter scenario occurs when there is a dense wind that becomes optically thick to the $\sim250$\,eV X-rays that ionize the inner shell of Ar and argued that this explained the case of T Cha \citep{bajaj2024}. RW\,Aur\,B would align with this case, consistent with the picture given by its [Ne III]/[Ne II]. Conversely, the ratio for VW\,Cha\,BC exceeds both of those limits, which could potentially imply an even denser wind \citep{sellek2024}.

\subsection{Variability in the MIR features of binary systems}

The three binary systems, RW\,Aur, VW\,Cha, and WX\,Cha, are active and variable accretors \citep{lisse2022, zsidi2022, fiorellino2022}, which could be related to the companions producing dynamical perturbations to the material, and increasing or decreasing the accretion rate as a part of the disk truncation process. The comparison of MIRI-MRS with IRS reveals a large variability of the continuum of RW\,Aur, and also in the line emission of VW\,Cha. WX\,Cha seems consistent both in continuum and lines. More interestingly, the line emission in RW\,Aur does not show obvious changes between the two epochs, and neither does the continuum of VW\,Cha, potentially hinting to the mechanism causing the variability.

In RW\,Aur\,A, the variability of the continuum has been attributed to a change in temperature of a very compact inner disk feature ($<0.1\,$au), potentially associated to a thermal instability of the innermost region of the accretion disk \citep{lisse2022}, thus moving the whole mid-infrared continuum up or down. Interestingly, this large continuum variability does not translate into a large line variability when comparing IRS to MIRI-MRS, potentially because most of the line flux is originating in radii than 0.1au. It is important to consider that the IRS observation was taken before RW\,Aur\,A started showing evidence of deep dimming events in 2010 \citep{rodriguez2013}, which continued irregularly for about a decade \citep[][]{petrov2015, facchini2016, koutoulaki2019}. Both mid-infrared observations, Spitzer and JWST, were taken before and after those dimming events when the star was in a quiescent stage, respectively, and thus RW\,Aur\,A could have had molecular line emission variability during this period.

Considering the disk of RW\,Aur\,A shows evidence of being warped, most likely due to the last dynamical interaction with RW\,Aur\,B \citep{kurtovic2025a}, the non-detection of line emission variability over the 15\,yr period also sets a constraint over the rate of illumination change due to the warp traveling through the disk. Depending on the disk properties, a warp can dissipate in a few outer disk orbits (about 300\,yrs for RW\,Aur\,A, \citet{Rowther2022}), or for timescales as long as $10^4$\,yrs \citep{kimmig2024}. Follow up observations with MRS could prove a better test of the long timescale variability of the molecular emission of RW\,Aur\,A, and provide constraints to the speed of the warp wave.

The variation of the line emission in VW\,Cha\,A is consistent with a change in emitting area, as a constant scale down by 0.5 of the line amplitude from MIRI-MRS is enough to match the IRS spectrum, thus hinting to an increase of emitting area by a factor of two. This change in emitting area does not exclude possible changes in temperature or column density of the emitting components, but those are more challenging to trace with the IRS spectral resolution. The increase of emitting area is not uniform across the different H$_2$O components, as the emission lines coming from cold H$_2$O are more than two times brighter in MIRI-MRS than IRS. 
The increase in emitting area could be related to changes in the accretion luminosity of VW\,Cha\,A, which could be warming the gas to further distances from the star. Such mechanism has also been proposed in another variable source, EX\,Lup \citep{smith2025}. 
If this was the case, then the line emission could change in timescales similar to that of accretion variability, which for VW\,Cha can change up to a factor of 3 within days \citep{fiorellino2022}.

The large variability in the [Ne II] emission of VW\,Cha remains to be explored. With IRS, the VW\,Cha system showed a brighter [Ne II] consistent with coming from a high velocity component, as subtracting the low-velocity component of [Ne II] detected in VW\,Cha\,BC produces almost no noticeable change to the IRS [Ne II] signature (see Fig.~\ref{fig:vwcha:comp-spt-jwst}). This high velocity component dramatically decreased in brightness between 2008 and 2023, and it is the only line emission that was brighter with IRS than with MIRI-MRS.

\subsection{The H$_2$ extended emission of the binaries}

We observe a bowl-like emission morphology in the extended H$_2$ of RW\,Aur, VW\,Cha, and WX\,Cha. This emission is similar in shape to that observed in SYCha \citep{schwarz2025}, suggesting it might originate from a wind. This interpretation is consistent with the ALMA observations of RW\,Aur, as the bowl has the same position angle as the outer disk in 1.3\,mm dust continuum \citep{kurtovic2025a}. In VW\,Cha and WX\,Cha, however, the lack of high angular resolution ALMA observations prevents us from relating the H$_2$ emission to disk structures or binary interactions.  

During close encounters, disk-companion interactions can launch material into unbound orbits, creating large spiral arms, and truncating the outer disks of each star. This is the case for RW\,Aur, where a fly-by about 300\,yrs ago \citep{kurtovic2025a} launched a large and bright spiral arm, which was first detected by \citet{Cabrit2006}, and has been studied in detail with simulations and observations \citep[e.g., ]{Dai2015, rodriguez2018}. At ALMA wavelengths, this arc is detected in $^{12}$CO J=2-1 (230.538\,GHz), as shown in panel (e) of Fig.~\ref{fig:rwaur:gallery_v2}. With the morphology from ALMA, we can now interpret the extended emission from MIRI-MRS. The emission in H$_2$\,S(1), for example, seems to follow the same morphology as that from ALMA, suggesting the H$_2$\,S(1) emission is also tracing material ejected during the interaction. When moving to higher energy transitions, such as H$_2$\,S(3) to S(8), the morphology changes completely, and it becomes a bowl-like shaped emission (see FIg.~\ref{fig:rwaur_extended}), with the same position angle as the outer disk of RW\,Aur\,A, as measured with ALMA. From ALMA, we know the radial extent of each disk in RW\,Aur\,A (about 0.4'' each). The H$_2$ emission with bowl-like shape is emitting farther than the size of the disks, meaning this H$_2$ is probably not gravitationally bound to RW\,Aur\,A or B. 

In RW\,Aur, the temperature of H$_2$ that reproduce the emission from S(5) to S(8) is higher over the same axis of the jet emission, probably tracing an outflow cavity, as it also has a distinct central velocity of emission (see panel j in Fig.~\ref{fig:h2model_highS}). The bowl-like emission shows the opposite trend, with lower temperature and higher column density. Thus, even though the emission seems spatially coherent in the sky plane, the H$_2$ might be coming from two different components. Other works have also seen H$_2$ in outflows with the same position angle as jet emission \citep[e.g.,][]{tychoniec2024, pascucci2025}, which might be the case in here.

\section{Conclusion} \label{sec:conclusions}

We have studied the new JWST/MIRI-MRS observations of three Class II binary systems, VW\,Cha, WX\,Cha, and RW\,Aur, as well as the extended emission of DF\,Tau. We have also compared our JWST/MIRI-MRS observations to Spitzer/IRS and ALMA, allowing us to analyze the systems over the time-domain, as well as interpreting the emission in the context of the outer disk and surroundings. We summarize our main findings as follows:

\begin{itemize}
    \item All the primary disks have a H$_2$O rich emission spectrum. Even though some of the secondary stars have a similar mass when compared to the primary star of their system, all of the secondary disks are mostly line poor to the sensitivity of our observations, with only RW\,Aur\,B showing signatures of the typically detected molecules: HCN, CO$_2$, and H$_2$O. 

    \item The water emission in the primary disks are well described by a collection of slab models, as in previous works \citep[e.g.,][]{temmink2024b, romeromirza2024}. In RW\,Aur, we find tentative evidence of a more structured water emission, which deviates from a simple power law in column density. The emitting area for the hottest water slab is consistent with the characteristic emitting radius of the CO, as observed in the M-band. 
    
    \item The primary and secondary disks are all affected by disk truncation from their companions, and despite having similar stellar mass ratios (except for WX\,Cha), the large difference in their mid-infrared spectra suggests that additional mechanisms to the disk size are dominating the inner disk chemistry, potentially disk substructures \citep[e.g.,][]{vlasblom2024}. 
    
    \item When comparing Spitzer/IRS to JWST/MIRI-MRS, a large variability in the molecular line emission is observed in VW\,Cha. This variability is consistent with a $50\%$ smaller emitting area in 2008 compared to 2023, probably originated from a change in the accretion luminosity \citep{zsidi2022}. No significant line variability is observed in the other sources. 

    \item The same comparison between IRS with MIRI-MRS finds a continuum variability for RW\,Aur. This variability seems to originate in the inner disk of RW\,Aur\,A, and it might be related to other variability events observed in the optical and NIR \citep{lisse2022}. 

    \item A high-velocity component [Ne II] emission is detected from all the primary disks, consistent with a jet-like outflow. The jets are spatially resolved in RW\,Aur, WX\,Cha, and DF\,Tau, with the last one showing tentative evidence of a variable PA$_{\text{jet}}$ as a function of distance. The secondaries VW\,Cha\,BC and RW\,Aur\,B have detections of [Ne II], [Ne III], and [Ar II], all of them consistent with a low-velocity component. These detections suggest a strong ionizing field, with wither EUV and X-ray dominated fields. 

    \item We are able to recover spatially resolved temperature and column density maps for the extended H$_2$ emission of RW\,Aur and VW\,Cha. In RW\,Aur, the low energy transition H$_2$\,S(1) shows a remarkable resemblance to the morphology of the cold $^{12}$CO from ALMA, suggesting that H$_2$ S(1) is also tracing material ejected during the last dynamical interaction.

\end{itemize}

As binaries can dynamically perturb the disk of their companions, disk misalignments, variably accretion and variable illumination can contribute to the mid infrared signatures we detect with MIRI-MRS. The depth to which we can interpret these features is deeper when observations from additional facilities are available, such as in RW\,Aur and DF\,Tau, where multiple high angular resolution observations with ALMA can aid in understanding the binary orbit, outer disks properties, and the environment of the systems. For VW\,Cha and WX\,Cha, where high-angular resolution ALMA data is still lacking, the interpretation is limited to results from optical and NIR studies, mostly on their variable accretion and spectral types. This work is another example for the need of multi-epoch and multi-wavelength observations to complement the spectra and extended emission from JWST/MIRI-MRS.

\section*{Acknowledgments}

We thank the anonymous referee for the constructive report and comments. We also thank Myriam Benisty and Stefano Facchini for the useful discussions. This work has been funded by the Deutsche Forschungsgemeinschaft (DFG, German Research Foundation) - 325594231, FOR 2634/2. E.v.D. acknowledges support from the ERC grant 101019751 MOLDISK and the Danish National Research Foundation through the Center of Excellence ``InterCat'' (DNRF150). M.T. and M.V. acknowledge support from the ERC grant 101019751 MOLDISK. T.H. and K.S. acknowledge support from the ERC Advanced Grant Origins 83 24 28. I.K. and E.v.D. acknowledge support from grant TOP-1614.001.751 from the Dutch Research Council (NWO). I.K. acknowledges funding from H2020-MSCA-ITN-2019, grant no. 860470 (CHAMELEON). V.C. acknowledges funding from the Belgian F.R.S.-FNRS. D.G. would like to thank the Research Foundation Flanders for co-financing the present research (grant number V435622N) and the European Space Agency (ESA) and the Belgian Federal Science Policy Office (BELSPO) for their support in the framework of the PRODEX Programme. T.K. acknowledges support from STFC Grant ST/Y002415/1. G.P. gratefully acknowledges support from the Max Planck Society. LS has received funding from the European Research Council (ERC) under the European Union’s Horizon 2020 research and innovation programme (PROTOPLANETS, grant agreement No. 101002188).

This work is based in part on observations made with the NASA/ESA/CSA James Webb Space Telescope. The data were obtained from the Mikulski Archive for Space Telescopes at the Space Telescope Science Institute, which is operated by the Association of Universities for Research in Astronomy, Inc., under NASA contract NAS 5-03127 for JWST. These observations are associated with program \#1282. The following National and International Funding Agencies funded and supported the MIRI development: NASA; ESA; Belgian Science Policy Office (BELSPO); Centre Nationale d’Etudes Spatiales (CNES); Danish National Space Centre; Deutsches Zentrum fur Luft- und Raumfahrt (DLR); Enterprise Ireland; Ministerio De Econom\'ia y Competividad; Netherlands Research School for Astronomy (NOVA); Netherlands Organisation for Scientific Research (NWO); Science and Technology Facilities Council; Swiss Space Office; Swedish National Space Agency; and UK Space Agency.

This work makes use of the following ALMA data: \\ 
ADS/JAO.ALMA\#2015.1.01506.S, \\ 
ADS/JAO.ALMA\#2016.1.00877.S, \\ 
ADS/JAO.ALMA\#2016.1.01164.S, \\ 
ADS/JAO.ALMA\#2017.1.01631.S, \\ 
ADS/JAO.ALMA\#2018.1.00973.S. \\ 
ADS/JAO.ALMA\#2019.1.01739.S. \\ 
ADS/JAO.ALMA\#2021.1.00854.S. \\ 
ALMA is a partnership of ESO (representing its member states), NSF (USA) and NINS (Japan), together with NRC (Canada), MOST and ASIAA (Taiwan), and KASI (Republic of Korea), in cooperation with the Republic of Chile. The Joint ALMA Observatory is operated by ESO, AUI/NRAO and NAOJ. 

This work also made use of data from the Combined Atlas of Sources with Spitzer IRS Spectra (CASSIS), a product of the IRS instrument team, supported by NASA and JPL. CASSIS is supported by the ``Programme National de Physique Stellaire'' (PNPS) of CNRS/INSU co-funded by CEA and CNES and through the ``Programme National Physique et Chimie du Milieu Interstellaire'' (PCMI) of CNRS/INSU with INC/INP co-funded by CEA and CNES. 
This work is in part based on observations made with ESO Telescopes at the Paranal Observatory under programme 110.244F (PI: Banzatti).

\bibliographystyle{aa.bst}
\bibliography{ms.bib}

\appendix

\section{Extracting individual spectra}\label{sec:app:extract-spectra}

\begin{figure*}[t]
  \centering
    \includegraphics[width=12cm]{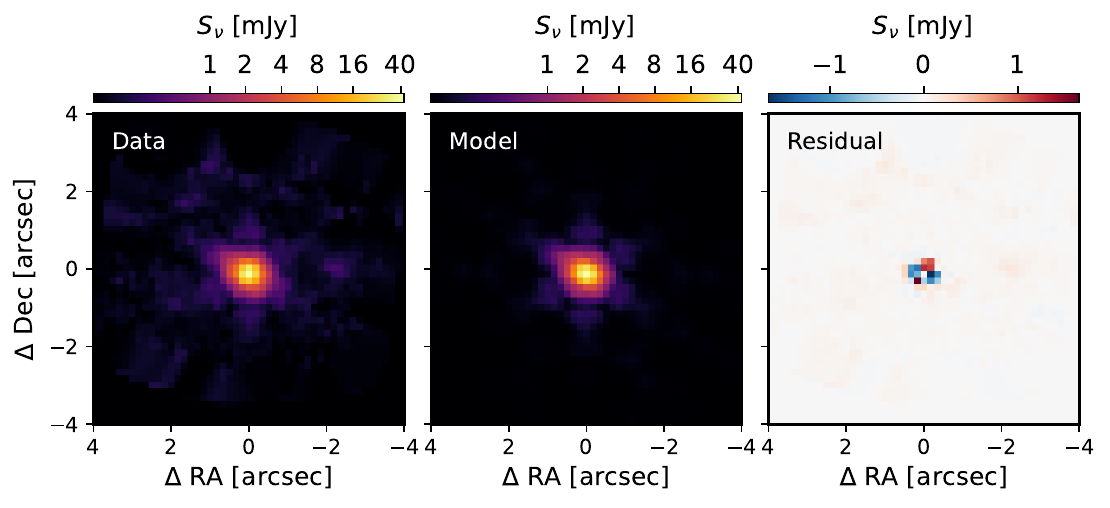}\\
    \vspace{-0.1cm}
    \caption{Example of forward model fitting in band 2A, channel 609, for WX\,Cha. The residuals are structured within the first FWHM, which is later added to the spectrum of WX\,Cha\,A with aperture photometry.  }
    \label{fig:example_psf_fit}
\end{figure*}

The peak position of each binary component is spatially separated in the MIRI-MRS detector, but the wings of the PSF overlap with each other. Thus, extracting the spectra of a source with aperture photometry would inevitably include flux coming from both its companion, contaminating the individual spectrum of each. In order to separate the emission of primary and secondary components, we forward model the emission of each source by fitting a theoretical PSF to each channel image in all the band cubes. The assumption behind this fit is that the flux contribution of each object is a point source. After fitting the PSF, we subtract the best model from the original image. The model contains almost all the flux of the source, but deviations between an ideal and empirical PSF, particularly in the first FWHM region, leave structured residuals with the shape of the undersampling curve \citep{argyriou2023, law2023}. Subtracting the PSF alleviates the problems related to the PSF-wing flux contamination, but leaves structured residual artifacts in the 1D spectra, as exampled in Fig.~\ref{fig:example_psf_fit}. Thus, after subtracting the best PSF model, we apply aperture photometry to the residuals, and later add that residual flux to the one obtained from the forward model. This process is repeated for every channel image in every band.

\section{Slab model results}

\begin{figure*}[t]
  \centering
    \includegraphics[width=15cm]{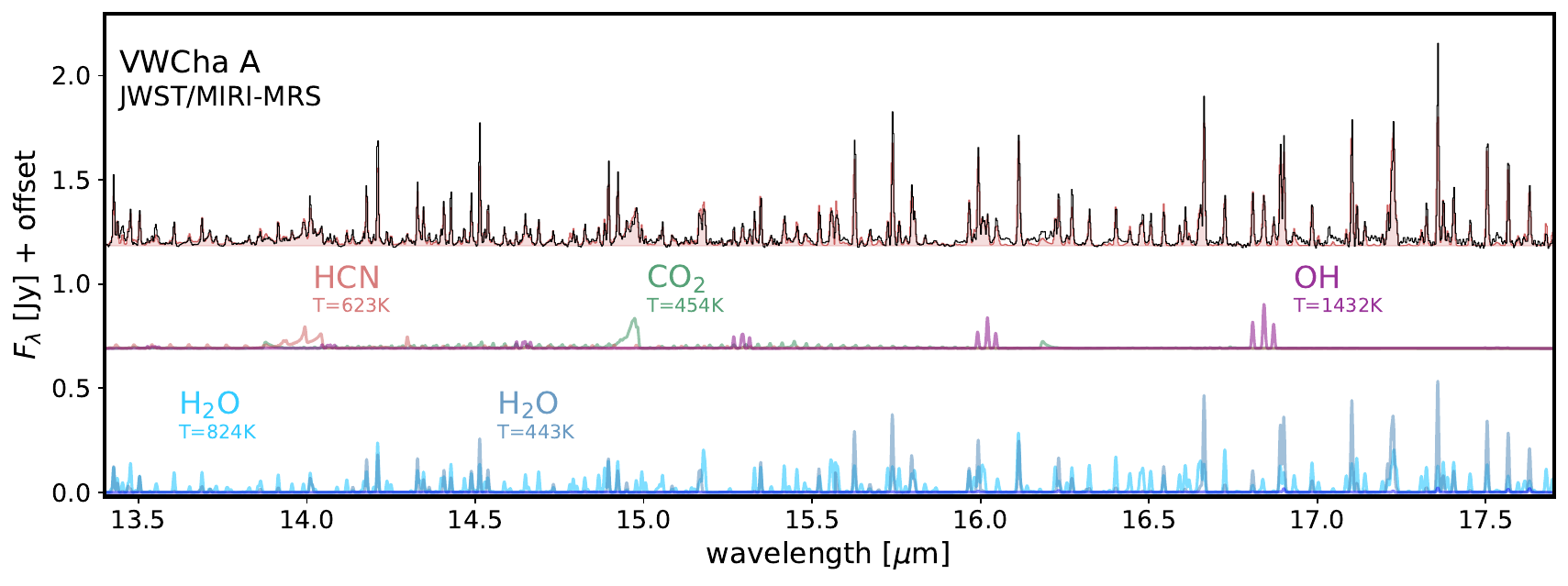}\\
    \includegraphics[width=15cm]{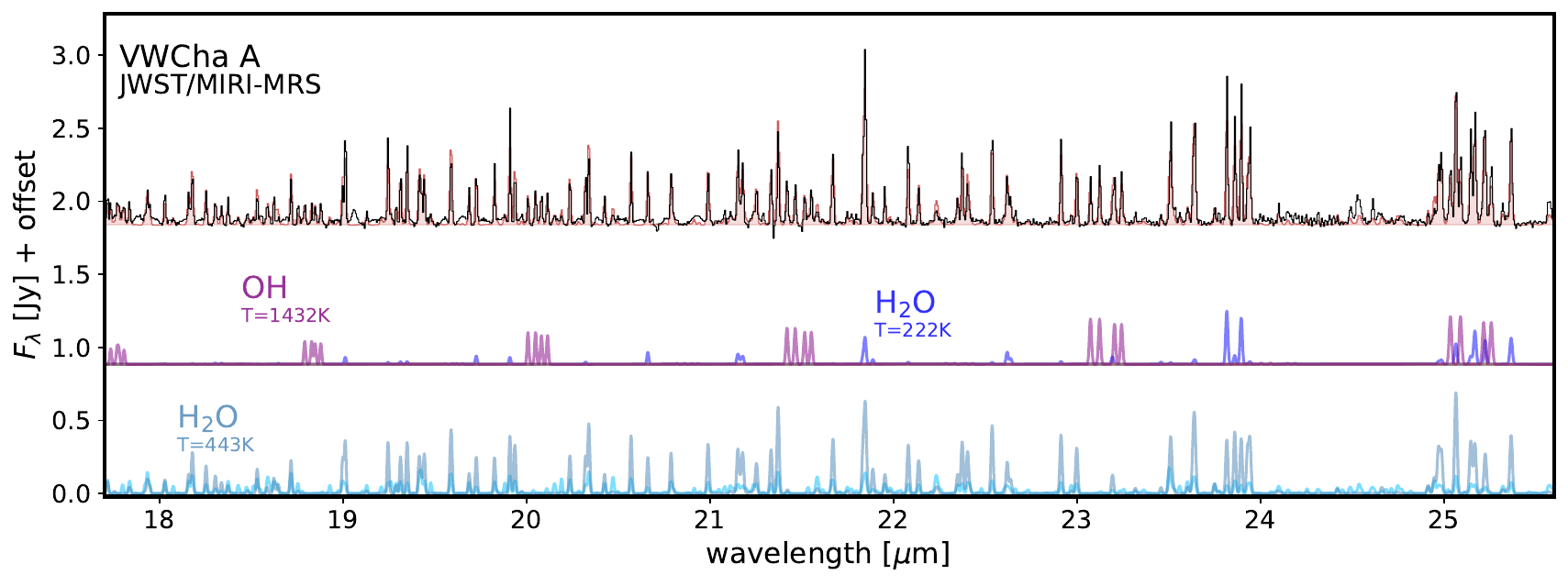}\\
    \vspace{-0.1cm}
    \caption{As in Fig.~\ref{fig:rwaur:spec-model}, but for VW\,Cha. }
    \label{fig:vwcha_specmodel}
\end{figure*}

\begin{figure*}[t]
  \centering
    \includegraphics[width=15cm]{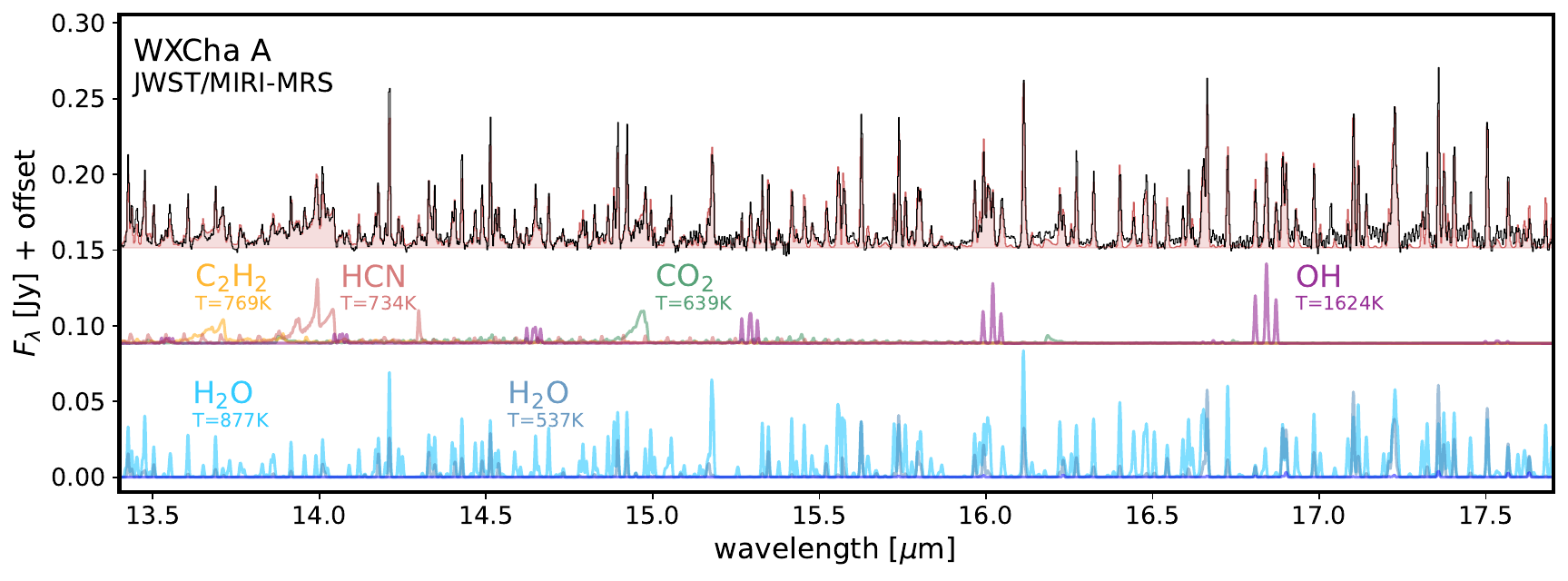}\\
    \includegraphics[width=15cm]{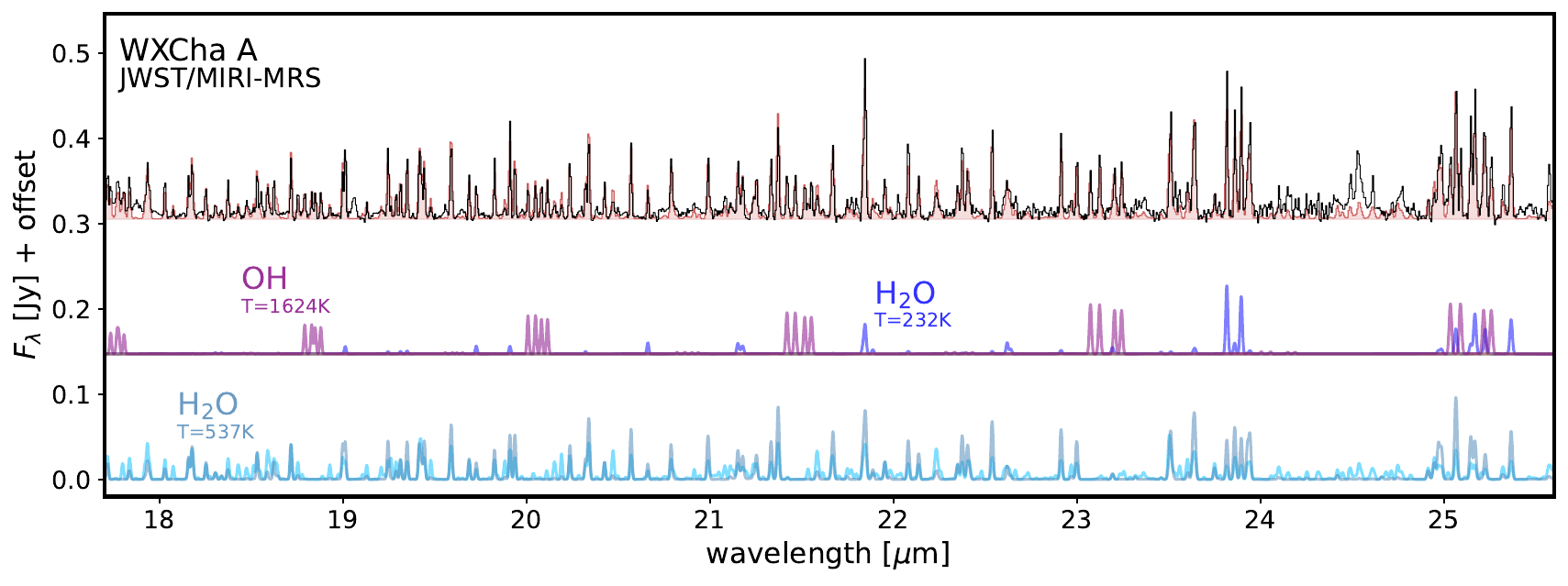}\\
    \vspace{-0.1cm}
    \caption{As in Fig.~\ref{fig:rwaur:spec-model}, but for WX\,Cha. }
    \label{fig:wxcha_specmodel}
\end{figure*}

In Tab.~\ref{tab:mcmc-slabs}, we present the results of the slab models for the molecular line emission in the MIRI-MRS bands 3B to 4C. This results are also described in Sect.~\ref{sec:res:slabs:water} and \ref{sec:res:slabs}. The Figures \ref{fig:vwcha_specmodel} and \ref{fig:wxcha_specmodel} show the maximum likelihood slab models for VW\,Cha and WX\,Cha, respectively.

\begin{table*}[t]
\centering
\caption{ \centering Maximum likelihood results for the slab models. }
\begin{tabular}{ c|c|c|c|c|c|c } 
  \hline
  \hline
\noalign{\smallskip}
Slab & Parameter & VW\,Cha\,A & WX\,Cha\,A & RW\,Aur\,A & $^\star$RW\,Aur\,B & units \\
\noalign{\smallskip}
  \hline
\noalign{\smallskip}
H$_2$O $\#1$ & $T$ &    $825 \pm 6$ &   $877 \pm 16$ & $1299 \pm 3$     &   450 & K \\
             & $R$ &  $0.41\pm0.01$ &  $0.22\pm0.01$ & $0.07 \pm 0.003$ &  0.45 & au  \\
             & $N$ & $19.61\pm0.02$ & $18.50\pm0.05$ & $19.99 \pm 0.01$ &  17.5 & $\log_{10}$\,cm$^{-2}$ \\
\noalign{\smallskip}
H$_2$O $\#2$ & $T$ &    $444 \pm 2$ &   $538 \pm 12$ &    $656.3 \pm 4$ &  -    & K \\
             & $R$ &  $1.74\pm0.01$ &  $0.59\pm0.02$ &  $0.41 \pm 0.01$ &  -    & au  \\
             & $N$ & $18.28\pm0.01$ & $17.82\pm0.05$ & $19.02 \pm 0.02$ &  -    & $\log_{10}$\,cm$^{-2}$ \\
\noalign{\smallskip}
H$_2$O $\#3$ & $T$ &    $222 \pm 2$ &   $232 \pm 11$ &  $333.7 \pm 1.5$ & -     & K \\
             & $R$ &  $5.99\pm0.03$ &  $6.88\pm2.82$ &  $1.35 \pm 0.01$ & -     & au \\
             & $N$ & $17.32\pm0.03$ & $16.32\pm0.55$ &  $19.16\pm 0.02$ & -     & $\log_{10}$\,cm$^{-2}$ \\
\noalign{\smallskip}
H$_2$O $\#4$ & $T$ & -              & -              &    $189.0 \pm 8$ & -     & K \\
             & $R$ & -              & -              &  $9.03 \pm 0.12$ & -     & au \\
             & $N$ & -              & -              &  $16.27\pm 0.13$ & -     & $\log_{10}$\,cm$^{-2}$ \\
\noalign{\smallskip}
C$_2$H$_2$   & $T$ &   $918 \pm 85$ &    $770\pm220$ &   $375.5 \pm 21$ & -     & K \\
             & $R$ & $<0.04\pm0.01$ &  $0.11\pm0.05$ &  $0.16 \pm 0.02$ & -     & au \\
             & $N$ & $17.99\pm0.01$ &  $17.48\pm1.6$ & $17.99 \pm 0.02$ & -     & $\log_{10}$\,cm$^{-2}$ \\
\noalign{\smallskip}
HCN          & $T$ &   $623 \pm 24$ &    $734\pm 48$ &   $540.9 \pm 15$ &   300 & K \\
             & $R$ &  $0.41\pm0.01$ &  $0.47\pm0.39$ &  $0.55 \pm 0.27$ &  0.12 & au \\
             & $N$ & $17.09\pm0.04$ & $16.14\pm0.58$ &   $16.4 \pm 0.4$ &  17.0 & $\log_{10}$\,cm$^{-2}$ \\
\noalign{\smallskip}
CO$_2$       & $T$ &   $455 \pm 12$ &    $640\pm158$ &     $593 \pm 14$ &   250 & K \\
             & $R$ &  $0.26\pm0.01$ &  $0.07\pm0.01$ &  $0.18 \pm 0.01$ &  0.18 & au \\
             & $N$ & $17.99\pm0.01$ & $17.99\pm0.01$ & $18.09 \pm 0.04$ &  17.0 & $\log_{10}$\,cm$^{-2}$ \\
\noalign{\smallskip}
OH           & $T$ &    $1432\pm 7$ &    $1624\pm42$ &    $1807 \pm 32$ & -     & K \\
             & $R$ &  $2.86\pm0.06$ &  $2.83\pm1.26$ &  $2.03 \pm 0.67$ & -     & au \\
             & $N$ & $15.00\pm0.02$ & $14.21\pm0.52$ & $14.54 \pm 0.37$ & -     & $\log_{10}$\,cm$^{-2}$ \\
\noalign{\smallskip}
  \hline
\noalign{\smallskip}
             & $RV$ &   $-13.5\pm 0.1$ &    $-11.1\pm0.6$ &    $14.0\pm0.2$ & -     & km s$^{-1}$ \\
\noalign{\smallskip}
  \hline
  \hline
\end{tabular}
\tablefoot{ \centering The values for RW\,Aur\,B were not fitted through MCMC. These uncertainties represent the 3$\sigma$ deviation in the distribution of each parameter. We warn to take the uncertainties as a lower limit, as any change deviation in the sensitivity from a pure thermal noise would not be captured by the MCMC, systematically underestimating the true uncertainty values. The radial velocity is measured relative to the rest velocity of the line emission. } 
\label{tab:mcmc-slabs}
\end{table*}

\section{Comparing IRS and MIRI}\label{sec:app:match-spt-jwst}

As the frequency resolution and frequency sampling of IRS and MIRI-MRS are not constant over their wavelength coverage, we divide the MIRI-MRS spectra on the individual bands from 3B to 4C to match them separately to the SH and LH bands from IRS. We do this empirically by finding the Gaussian kernel needed to match the spectra of both instruments. After obtaining the IRS continuum subtracted spectra, we ran an MCMC to find the maximum likelihood convolution width for each matching pair (3B-SH, 3C-SH, 4A-SH, 4A-LH, 4B-LH, 4C-LH). After the convolution, we match the sampling of the MIRI-MRS spectra to the IRS with \texttt{spectres} \citep{carnall2017}. 
For the spectra of RW\,Aur\,A and WX\,Cha\,A, we find consistent results for the convolution constants, which confirms that both have similar line emission between the two instruments. In VW\,Cha, these convolution constants deviate from the other two disks, as the line fluxes differ for each epoch. Thus, for VW\,Cha, we convolved the MIRI-MRS spectra using the constants obtained for RW\,Aur.

\begin{figure*}[t]
  \centering
    \includegraphics[width=18cm]{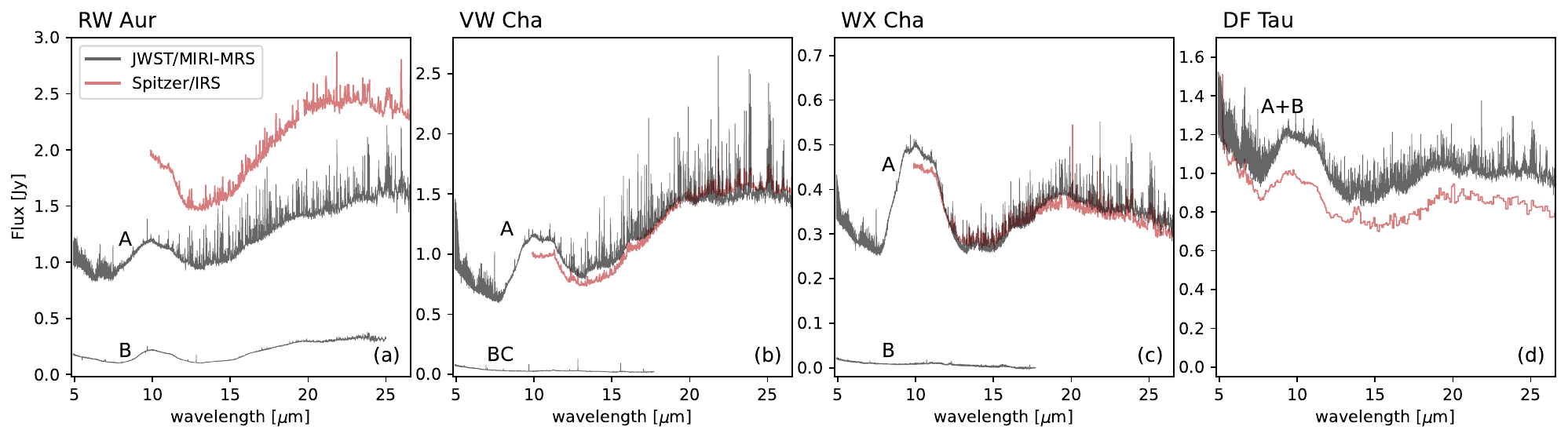}\\
    \vspace{-0.1cm}
    \caption{Comparison Spitzer/IRS and JWST/MIRI-MRS. }
    \label{fig:app:comp-spt-jwst}
\end{figure*}

\begin{figure*}[t]
  \centering
    \includegraphics[width=17cm]{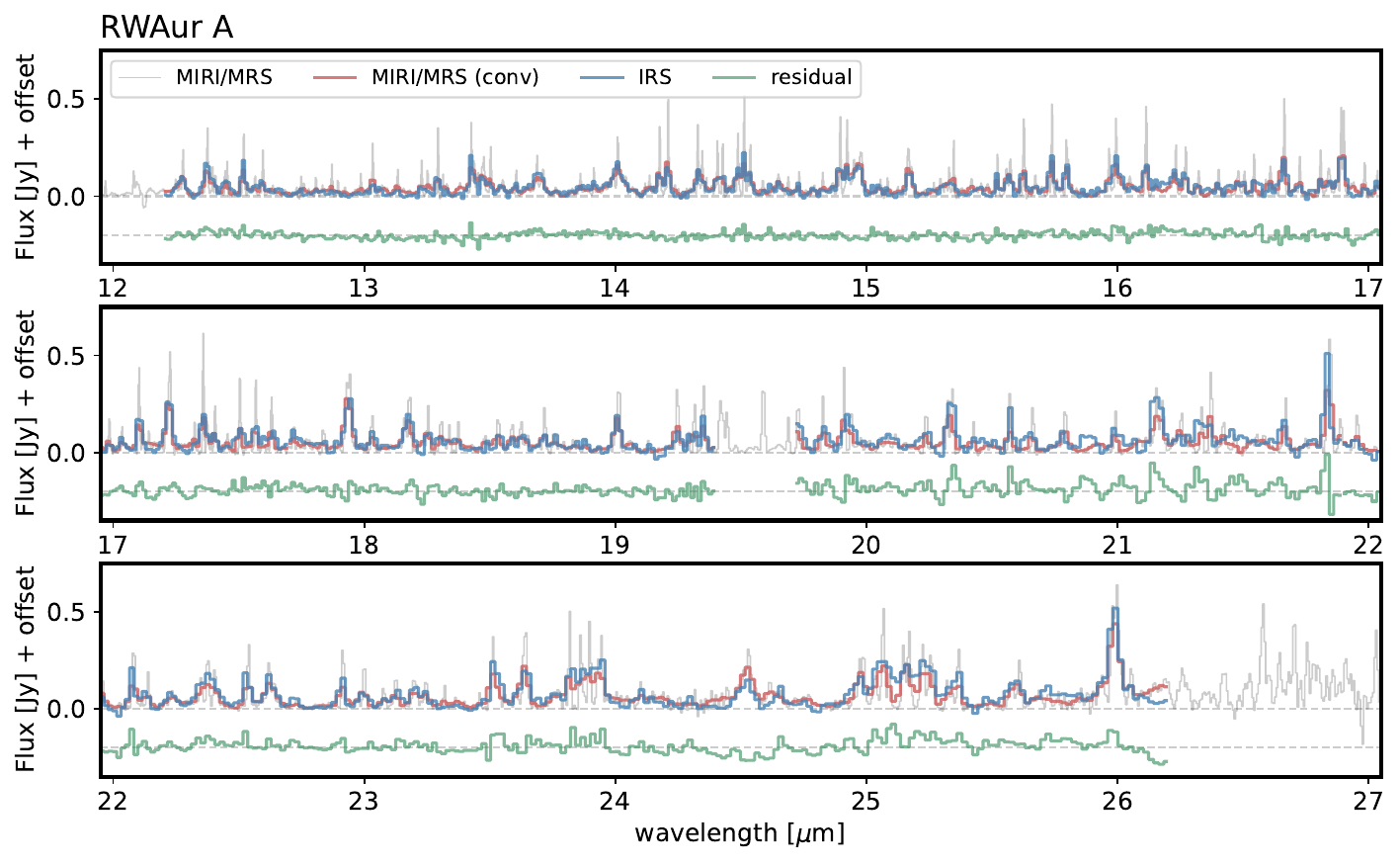}\\
    \vspace{-0.1cm}
    \caption{Comparison of MIRI-MRS to IRS, for RW\,Aur A. }
    \label{fig:app:rwaur:comp-spt-jwst}
\end{figure*}

\begin{figure*}[t]
  \centering
    \includegraphics[width=17cm]{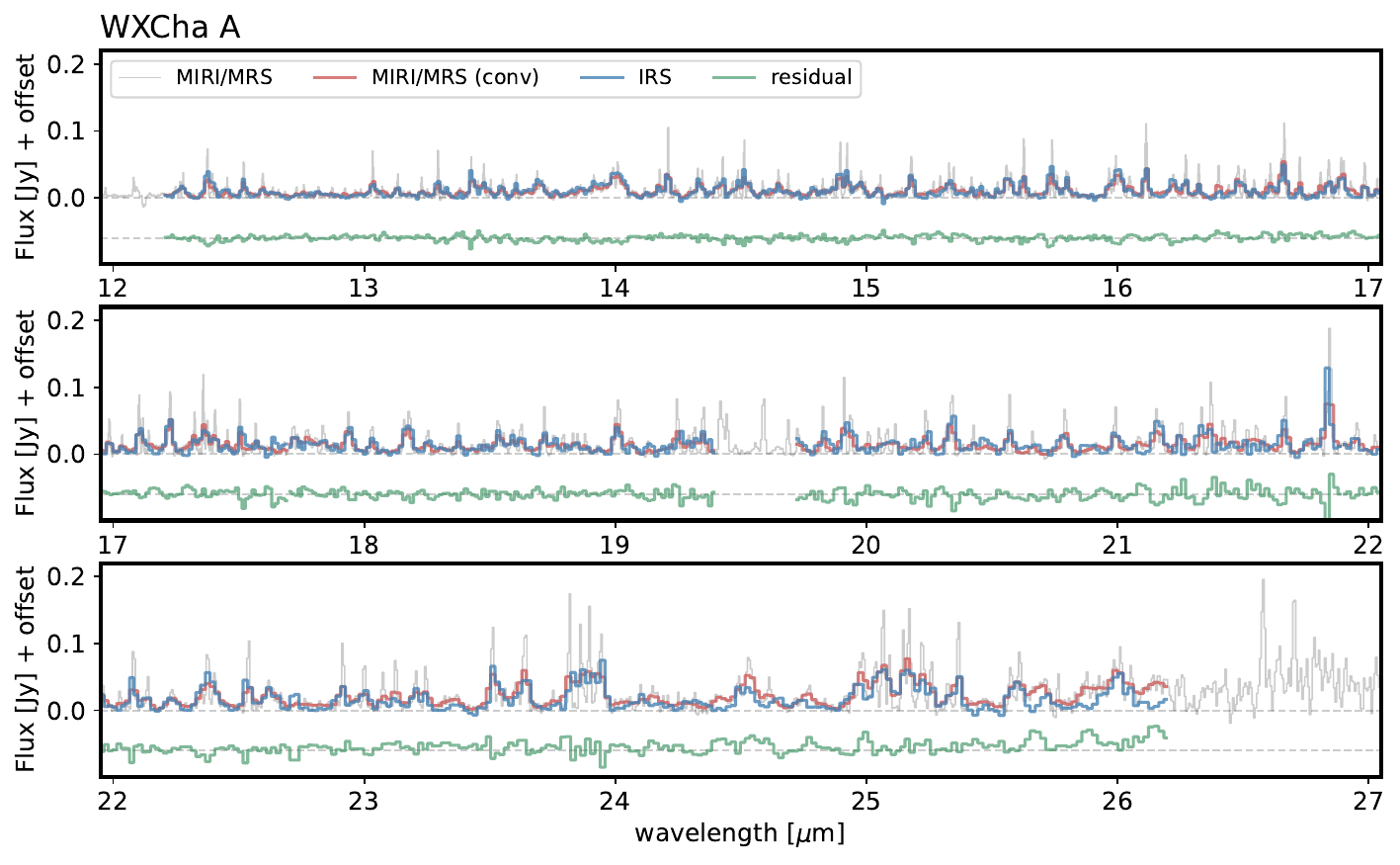}\\
    \vspace{-0.1cm}
    \caption{Comparison of MIRI-MRS to IRS, for WX\,Cha A. }
    \label{fig:app:wxcha:comp-spt-jwst}
\end{figure*}

\section{Fitting the noble gas emission} \label{sec:app:fit_noblegas}

We fit the noble gas emission with one or two Gaussian components, which have three free parameters: Peak brightness, line width, and radial velocity relative to the rest wavelength of each line. 
Each Gaussian is then convolved to the velocity resolution of MIRI-MRS at the observed wavelength, as measured by \citet{banzatti2025}. We run our fit by finding the combination of parameters with the maximum likelihood, using an MCMC with uniform priors for each parameter. These results are reported in Fig.~\ref{fig:all-neon-argon}, as well as in Tab.~\ref{tab:noble_gas}.

\begin{table*}[t]
\centering
\caption{ \centering Results of Gaussian fit to detected noble gas emission}
\begin{tabular}{ c|l|c|c } 
  \hline
  \hline
\noalign{\smallskip}
Source & Line & Flux & RV  \\
       &      & [$10^{-15}$ erg cm$^{-2}$ s$^{-1}$] & [km s$^{-1}$]  \\
\noalign{\smallskip}
  \hline
\noalign{\smallskip}
VW\,Cha\,A  & [Ne II]  (blue) & $8.04 \pm 0.27 $ & $-175 \pm 3 $ \\
            & [Ne II]  (rest) & $6.62 \pm 0.32 $ & $5 \pm 3 $ \\
\noalign{\smallskip}
  \hline
\noalign{\smallskip}
WX\,Cha\,A  & [Ne II]  (blue) & $1.87 \pm 0.18 $ & $-102 \pm 16 $ \\
            & [Ne II]  (red)  & $1.05 \pm 0.23 $ & $72 \pm 13 $ \\
\noalign{\smallskip}
  \hline
\noalign{\smallskip}
RW\,Aur\,A  & [Ne II]  (blue) & $11.96 \pm 0.18 $ & $-130 \pm 3 $ \\
            & [Ne II]  (rest) & $2.56  \pm 0.14 $ & $ 25 \pm 4 $ \\
\noalign{\smallskip}
  \hline
\noalign{\smallskip}
VW\,Cha\,BC & [Ne II]  (blue) & $1.19  \pm 0.09 $ & $ -163 \pm 7 $ \\
            & [Ne II]  (rest) & $8.67 \pm 0.09 $ & $ 8 \pm 1 $ \\
            & [Ne III] (rest) & $5.67 \pm 0.19 $ & $ 25 \pm 2 $ \\
            & [Ar II]  (rest) & $2.16  \pm 0.29 $ & $ 10 \pm 6 $ \\
\noalign{\smallskip}
  \hline
\noalign{\smallskip}
RW\,Aur\,B  & [Ne II]  (rest) & $6.35 \pm 0.14 $ & $-3 \pm 1 $ \\
            & [Ne III] (rest) & $0.87  \pm 0.15 $ & $8 \pm 17 $ \\
            & [Ar II]  (rest) & $2.51  \pm 0.39 $ & $-6 \pm 8 $ \\
\noalign{\smallskip}
  \hline
  \hline
\end{tabular}
\tablefoot{ \centering The uncertainty was measured from the $3\sigma$ dispersion of allowed solutions in the MCMC, which used Gaussian profiles to describe the emission. Any deviation from a single Gaussian component would contribute to artificially lower the uncertainty, and thus the true uncertainty of the measurement is underestimated in these cases. In each system, the radial velocity of the line center (RV) is measured relative to the central velocity of the spectra for the primary disks, as estimated from the slab modeling and presented in Table \ref{tab:mcmc-slabs}. } 
\label{tab:noble_gas}
\end{table*}

\section{CO emission from VLT/CRIRES+} \label{sec:app:criresp}

VW Cha and WX Cha were observed with VLT/CRIRES+ \citep{dorn2014, dorn2023} to obtain the CO fundamental line profiles. Both targets were observed in two filters, M4211 and M4368, to provide sufficient coverage of both the low- and high-J line transitions \citep[see][for a detailed description of the instrument filters in the L and B band]{grant2024a}. VW Cha was observed on April 16, 2023. WX Cha was observed in the night of December 30 to 31, 2022. VW Cha was observed for $\sim$16 minutes on source and WX Cha was observed for $\sim$40 minutes on source, and both were proceeded by an observation of a telluric standard star that was used to remove the telluric lines. A slit width of $0.4''$ was used in the observations and AO was not utilized due to the low declination/high airmasses of the targets. The reduction of these datasets was done with a customized pipeline, to account for the missing calibration lamps in the M-band \citep[see][for a discussion of wavelength shifts between the A and B nodding positions]{grant2023}. Our customized data reduction uses the telluric absorption lines for wavelength calibration, thus avoiding the need for cross-correlating the spectra of each nod. A detailed report on our customized pipeline will be published in Kurtovic et al., (in prep). The M-band spectrum of RW\,Aur\,A was obtained from the webpage spexodisks\footnote{\url{https://spexodisks.com}} \citep{wheeler2024}, which are VLT/CRIRES observations from October 15, 2007, and the details of this dataset can be found in \citet{brown2013}. The RW\,Aur\,A observations were obtained during a quiescent period for the optical brightness of this star, before it began its dimming events in 2010 \citep{rodriguez2013}.

In order to increase the S/N of the CO emission, we stacked the line profiles of VW\,Cha and WX\,Cha by combining the P-26, P-27, P-30, P-31, and P-32 transitions. For RW Aur, we only used the clean and confidently detected P-21 transition. More details on the procedure for line stacking are given in \citep{temmink2024}. The stacked spectra, which represents the high-J transitions, is analyzed by fitting two Gaussian profiles, allowing the fit of double-peaked profiles, as well as profiles described with broad and narrow velocity components \citep[see also][]{banzatti2022}. These Gaussians were fitted using the python-package \texttt{lmfit} \citep{newville2024}

\section{Fitting the extended H$_2$ emission}\label{sec:app:fit_h2}

As the position of the sources in the detector changes with wavelength, the Moment 0 map of each H$_2$ transition needs to be resampled to make sure that every pixel is covering the same region of the sky. We interpolate each Moment 0 map to a new supersampled image with pixel size of 10\,mas, which is about 40 times smaller than the full width at half maximum (FWHM) of the PSF. This supersampling is done with bilinear interpolation, which ensures flux conservation and does not distort the original flux morphology \citep[e.g.,][]{stevenson2012}. Then, we integrate the supersampled image down to pixels of 0.2'' in size, which is half the FWHM. The flux is checked in every step of the resampling, and we confirm the resampled image has less than 1\% of difference in total flux compared to the original image. The resampled H$_2$\,S(5) to S(8) transitions of RW\,Aur and VW\,Cha are shown in the upper rows of Fig.~\ref{fig:h2model_highS}.

The resampled images have a consistent pixel grid, which allows to model each pixel with a slab model across the different H$_2$ transitions. For each slab model, we integrate the total flux of each transitions between S(5) to S(8). The temperature of the slab ($T_{\text{H}_2}$) and the column density ($N_{\text{H}_2}$) are left as free parameters, and the emitting area ($A_{\text{H}_2}$) is fixed to the pixel area. 
We run an MCMC for every pixel, with 80 walkers and 500 steps, using an uniform prior for each parameter. The convergence is achieved in less than 100 steps, and the remaining 400 steps are a conservative MCMC length to ensure the stability of the minimum $\chi^2$ result. 
For every pixel, we save the $T_{\text{H}_2}$ and $N_{\text{H}_2}$ of the best solution, thus producing an image for the spatial distribution of these parameters. 
In pixels with high uncertainty and low flux contribution from H$_2$, the values for $T_{\text{H}_2}$ and $N_{\text{H}_2}$ can become unrealistic and deviate largely from the neighboring pixels, as the solution is not constrained at all in the absence of flux.
Thus, we apply a filter by excluding solutions with temperatures lower than 500\,K or higher than 2000\,K, as well as column densities below $10^{17}$ or higher than $10^{19}$ in units of cm$^{-2}$. These results are shown in the lower rows of \ref{fig:h2model_highS}.

\section{Molecular emission in RW\,Aur\,B}\label{sec:app:rwaurb_fit}

\begin{figure*}[t]
  \centering
    \includegraphics[width=17cm]{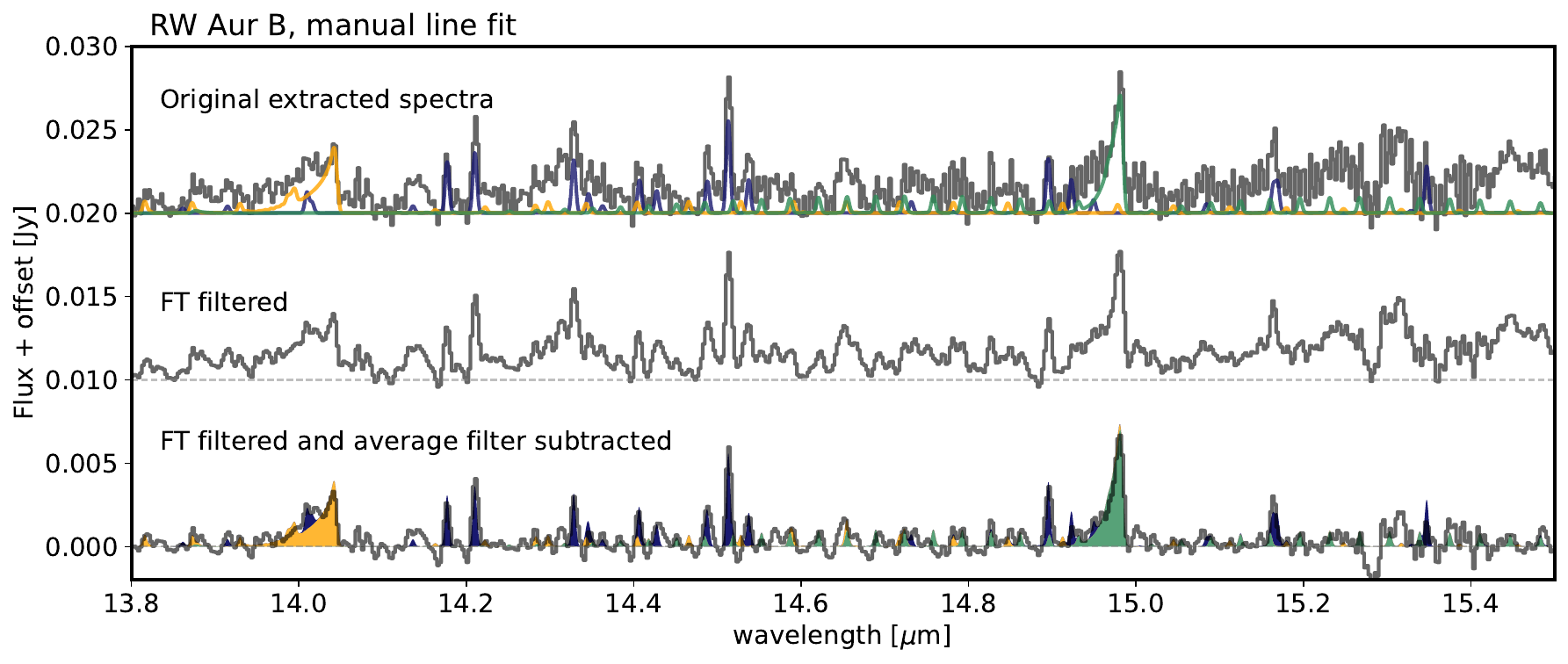}\\
    \vspace{-0.1cm}
    \caption{ The molecular line emission of RW\,Aur\,B. In the top is the original continuum subtracted spectra, with the same continuum subtraction algorithm used for the primary disks. In the middle is the spectra after filtering the signals with frequency smaller than than 0.5\,px$^{-1}$. In the bottom, the spectra after frequency filtering and subtracting a masked average with a window size of 10\,px. For reference, slab models for HCN, H$_2$O, and CO$_2$, are shown with the original and with the modified spectra. }
    \label{fig:app:rwaurb_manual_fit}
\end{figure*}

The sensitivity in the spectra of RW\,Aur\,B is limited by fringing and undersampling of the PSF, which results in low- and high- frequency oscilations in the spectra which are not originated in the source emission. Due to the faintness of the line emission, these artifacts have a similar amplitude to those of real molecular line emission, interfering with the recovery of the line emission properties. 

For the purpose of a better visualization of the molecular emission, we applied a high frequency mask to the spectrum of RW\,Aur\,B, removing any signal with a frequency equal or smaller than 1\,px$^{-1}$. After removing the potential line emission, we applied an average filter to calculate the lower frequency variations across the spectrum. In Fig.~\ref{fig:app:rwaurb_manual_fit}, we present the spectrum in the different stages of the process, also showing the line emission before and after applying the frequency and average filters. The molecular line emission was fitted manually by varying the parameters for temperature and column density, thus, they should only be taken for reference.

\section{Additional information on the extended emission}

All the binary systems studied in this work show evidence of extended emission, which we summarize in Tab.~\ref{app:tab:lines_extended}. For DF\,Tau, the extended emission is compared to high angular resolution observations from ALMA in Fig.~\ref{fig:app:dftau-alma-jwst}, where the $^{12}$CO $\nu=0$ $J=2-1$ is confidently detected around the primary disk, and tentatively around the secondary. The position angle of the red and blueshifted sides for the $^{12}$CO suggest a close to perpendicular orientation between DF\,Tau\,A circumstellar disk and the [Ne II] jet observed by MIRI-MRS. 
The extended emission in the different H$_2$ transitions from S(1) to S(8) are shown in Fig.~\ref{fig:rwaur_extended}, \ref{fig:vwcha_extended}, and \ref{fig:wxcha_extended}, for RW\,Aur, VW\,Cha, and WX\,Cha, respectively. For DF\,Tau, there is no confident detection of extended H$_2$ emission.

\begin{figure*}[t]
  \centering
    \includegraphics[width=18cm]{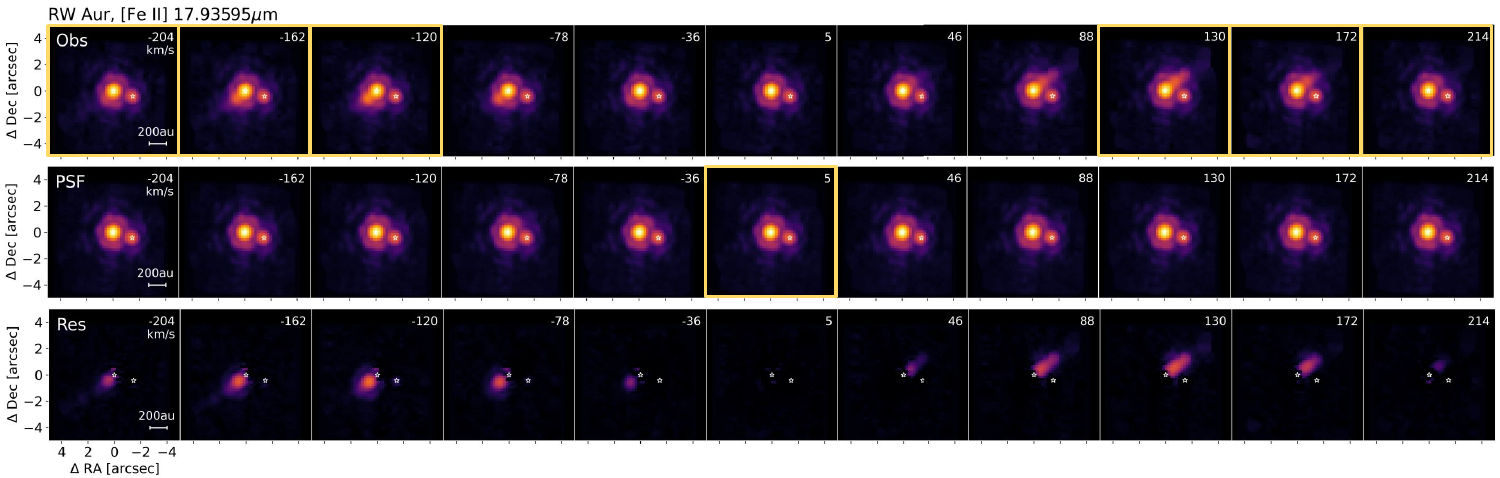}\\
    \vspace{-0.1cm}
    \caption{An example of PSF subtraction is shown from upper row to lower row, using the observation of [Fe II] of RW\,Aur as an example. The panels marked with color in the observation row show which channels were used to calculate the median PSF for the marked channel in the middle row. This empirical PSF is scaled to match the peak flux of the observation. When subtracting the scaled PSF, the residuals contain the extended emission for the specific line. }
    \label{app:fig:median_psf}
\end{figure*}

\begin{table*}[t]
\centering
\caption{ \centering Emission lines with confident detections of extended emission in each binary system. }
\begin{tabular}{ l|c|c|c|c|c|c } 
  \hline
  \hline
\noalign{\smallskip}
Line        & Freq.    & band & RW\,Aur  & VW\,Cha  & WX\,Cha  & DF\,Tau \\
            & [$\mu$m] &          &          &          &          &         \\
\noalign{\smallskip}
  \hline
\noalign{\smallskip}
 H$_2$ S(8)  & 5.05303  & 1A       & ext      & ext      &          &         \\
 H$_2$ S(7)  & 5.5112   & 1A       & ext      & ext      &          &         \\
 H$_2$ S(6)  & 6.1086   & 1B       & ext      & ext      &          &         \\
 H$_2$ S(5)  & 6.9095   & 1C       & ext      & ext      & ext      &         \\
 H$_2$ S(4)  & 8.0251   & 2A       & ext      & ext      & ext      &         \\
 H$_2$ S(3)  & 9.6649   & 2B       & ext      & ext      & ext      &         \\
 H$_2$ S(2)  & 12.2786  & 3A       & ext      & ext      & ext      &         \\
 H$_2$ S(1)  & 17.0348  & 3C       & ext      & ext      & ext      &         \\
\noalign{\smallskip}
  \hline
\noalign{\smallskip}
 Fe II       & 5.340169 & 1A       & jet      &          &          &         \\
             & 6.721283 & 1C       & jet      &          &          &         \\
             & 17.93595 & 3C       & jet      &          &          &         \\
             & 17.93595 & 4A       & jet      &          &          &         \\
             & 24.51925 & 4C       & jet      &          &          &         \\
             & 25.98829 & 4C       & jet      &          &          &         \\
\noalign{\smallskip}
  \hline
\noalign{\smallskip}
 Ni II       & 6.6360   & 1C       & jet      &          &          &         \\
             & 10.6822  & 2C       & jet      &          &          &         \\
             & 12.7288  & 3A       & jet      &          &          &         \\
\noalign{\smallskip}
  \hline
\noalign{\smallskip}
 Ne II     & 12.81355 & 3A       & jet+ext  & ext      & jet      & jet$+$ext \\
 Ne III    & 15.5551  & 3C       &          & ext      &          &          \\
 Ar II     & 6.985274 & 1C       & jet      & ext      &          & jet      \\
\noalign{\smallskip}
  \hline
  \hline
\end{tabular}
\tablefoot{ \centering The naming \textbf{jet} is for detection of a high-velocity component to the extended emission ($\approx100$\,km\,s$^{-1}$ from line center), and \textbf{ext} is for detection of extended emission, but at low velocity ($<50$\,km\,s$^{-1}$ from line center). VW\,Cha has a detection of a high velocity [Ne II] in its spectra, but it is not confirmed in the extended emission. }
\label{app:tab:lines_extended}
\end{table*}

\begin{figure*}[t]
  \centering
    \includegraphics[width=15cm]{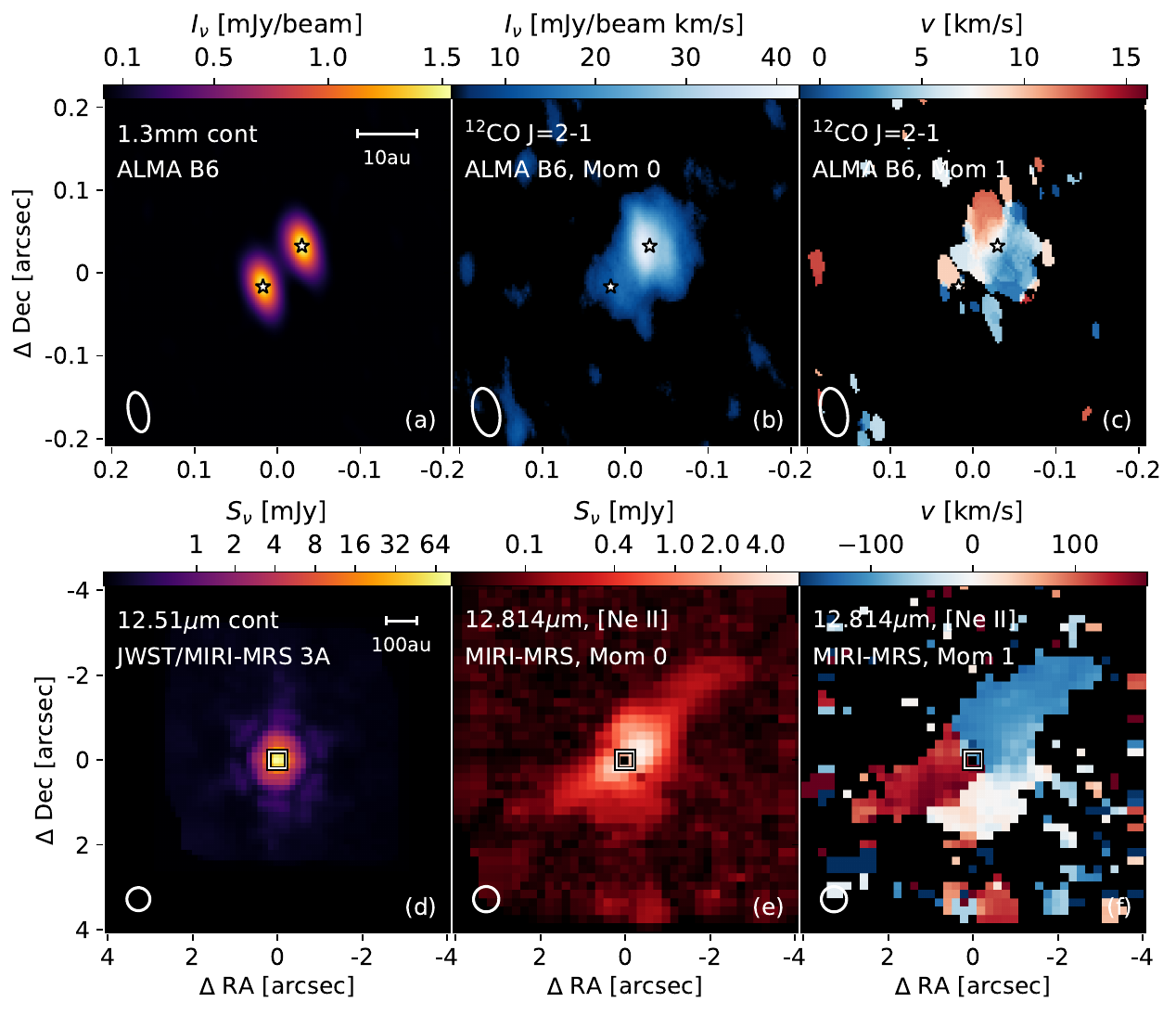}\\
    \vspace{-0.1cm}
    \caption{The upper panels show DF\,Tau as observed with ALMA, showing from left to right: The 1.3\,mm dust continuum emission, the $^{12}$CO J=2-1 Moment 0, and the $^{12}$CO J=2-1 moment 1. These datasets were presented in \citet{grant2024}. In the lower panels show the MIRI-MRS observations, from left to right: The 12.51$\mu$m continuum emission, the Moment 0 frequency of [Ne II], and the velocity at peak emission of [Ne II] relative to the rest frequency. The ellipse in the corner of each panel represents the FWHM of the PSF of each image. The box at the center of panels (d,e,f) shows the spatial extent of panels (a,b,c). }
    \label{fig:app:dftau-alma-jwst}
\end{figure*}

\begin{figure*}[t]
  \centering
    \includegraphics[width=17cm]{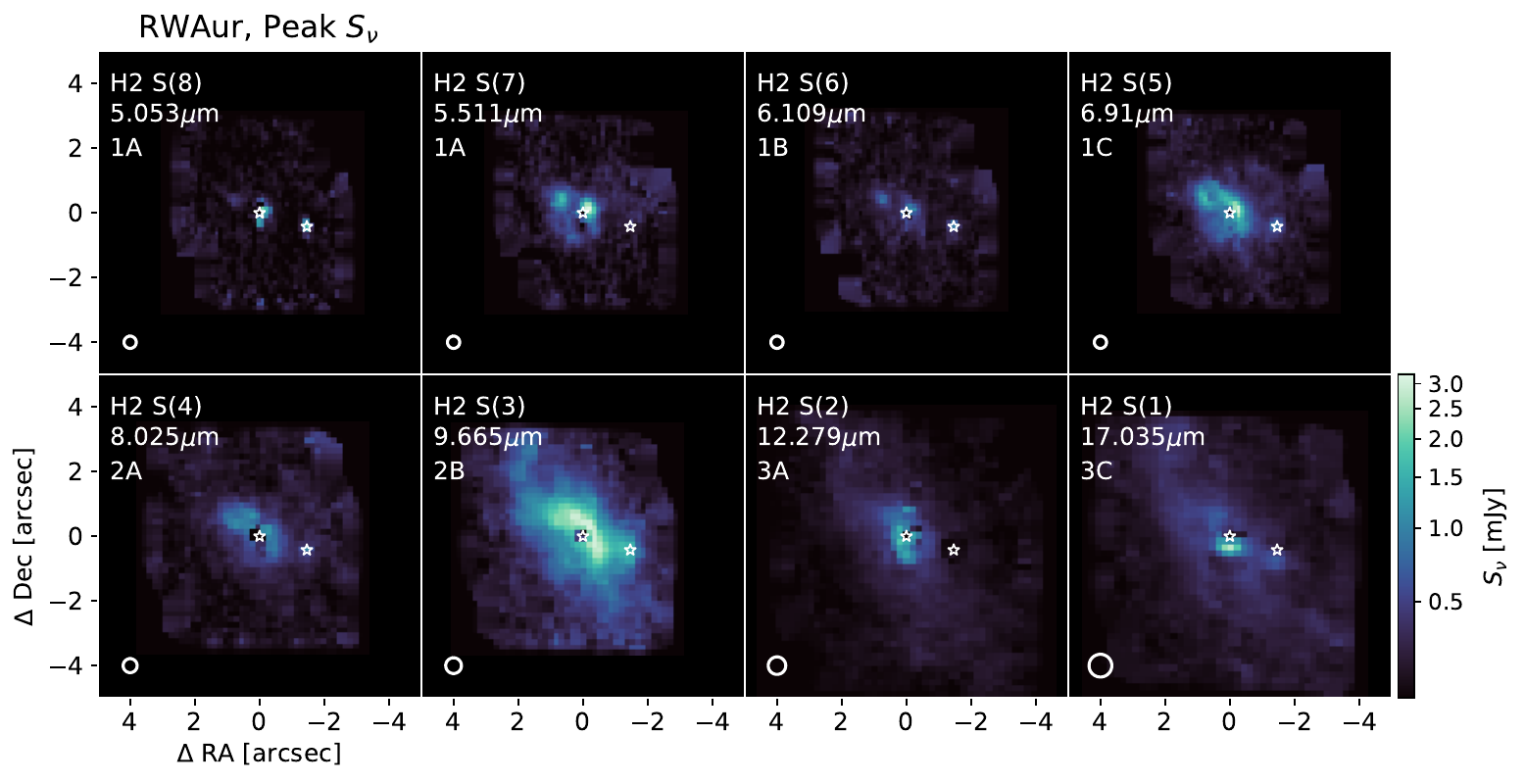}\\
    \vspace{-0.1cm}
    \caption{Peak brightness moment map for the extended emission of pure rotational H$_2$ emission, for the RW\,Aur system. }
    \label{fig:rwaur_extended}
\end{figure*}

\begin{figure*}[t]
  \centering
    \includegraphics[width=17cm]{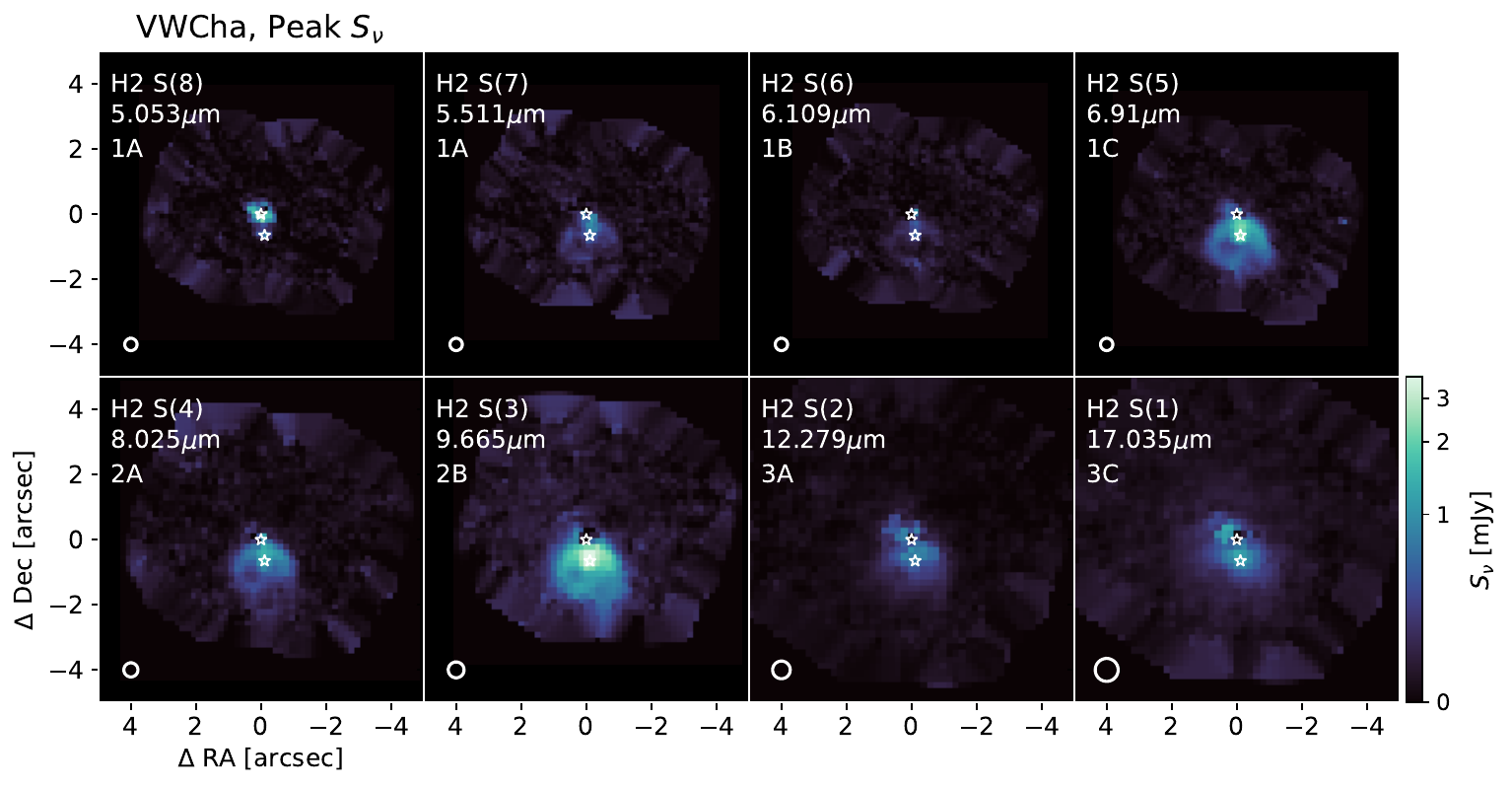}\\
    \vspace{-0.1cm}
    \caption{Peak brightness moment map for the extended emission of pure rotational H$_2$ emission, for the VW\,Cha system. }
    \label{fig:vwcha_extended}
\end{figure*}

\begin{figure*}[t]
  \centering
    \includegraphics[width=17cm]{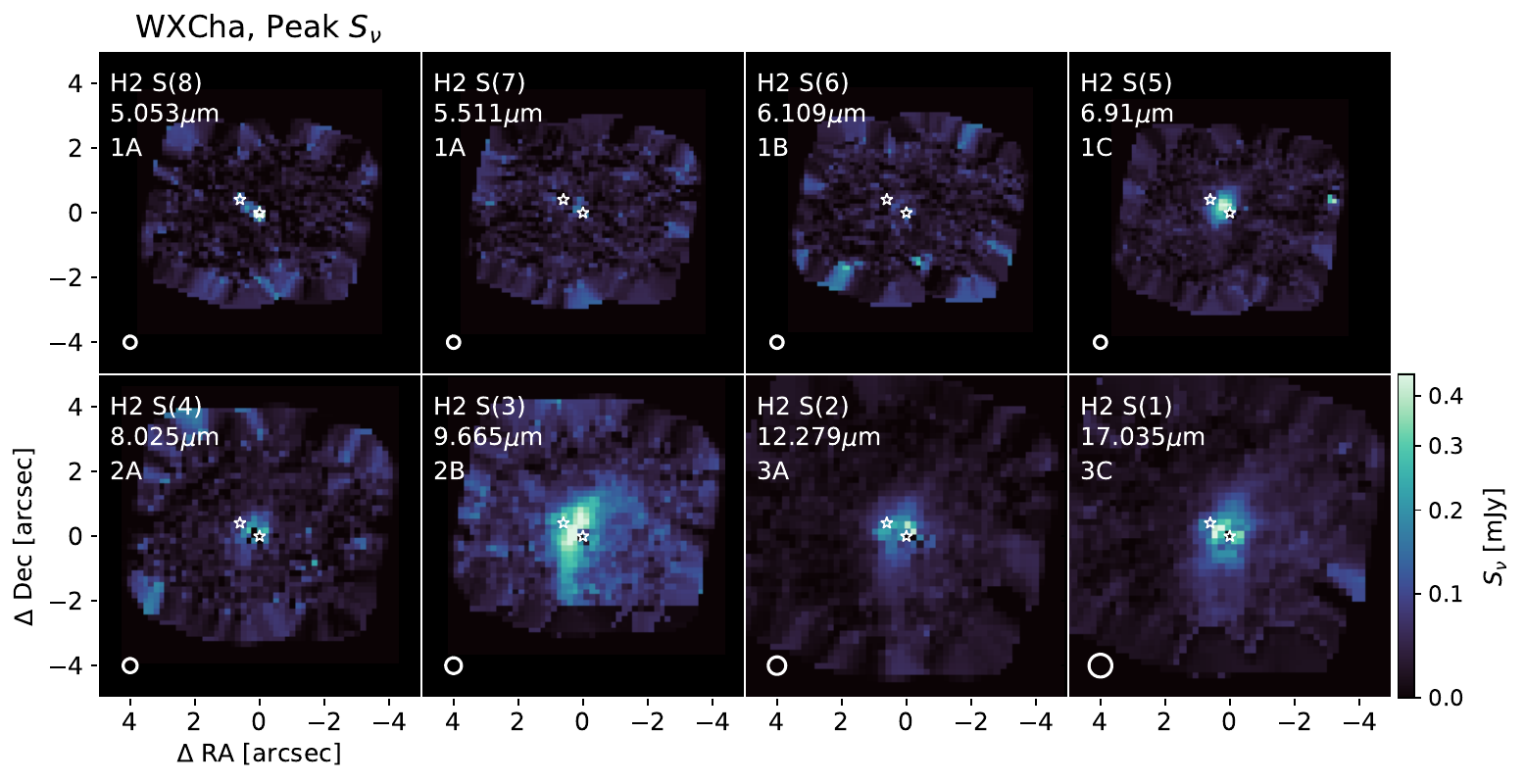}\\
    \vspace{-0.1cm}
    \caption{Peak brightness moment map for the extended emission of pure rotational H$_2$ emission, for the WX\,Cha system. }
    \label{fig:wxcha_extended}
\end{figure*}

\end{document}